\documentclass[11pt]{article}
\usepackage{microtype}
\usepackage{booktabs}
\usepackage{ifpdf}
\ifpdf
\usepackage[pdftex]{graphicx}
\else
\usepackage{graphicx}
\fi
\usepackage[sc,noBBpl]{mathpazo}
\usepackage[mathscr]{eucal}
\usepackage{amsmath,amsfonts,amssymb,amsthm}
\usepackage[square,sort]{natbib}
\theoremstyle{plain}
\newtheorem{theorem}{Theorem}
\newtheorem{lemma}[theorem]{Lemma}
\newtheorem{proposition}[theorem]{Proposition}

\newtheorem{conjecture}[theorem]{Conjecture}
\newtheorem{definition}[theorem]{Definition}
\newtheoremstyle{note}{\topsep}{\topsep}{\slshape}{}{\scshape}{}{ }{}
\theoremstyle{note}

\newtheorem{remark}[theorem]{Remark}
\numberwithin{equation}{section}
\numberwithin{theorem}{section}

\newcommand\efg{{\mathfrak g}}

\newcommand\cG{{\mathcal G}}

\newcommand\scA{{\mathscr A}}

\newcommand\scC{{\mathscr C}}
\newcommand\scD{{\mathscr D}}

\newcommand\scG{{\mathscr G}}

\newcommand\scI{{\mathscr I}}

\newcommand\scL{{\mathscr L}}
\newcommand\scM{{\mathscr M}}

\newcommand\scV{{\mathscr V}}

\newcommand\mvector{\boldsymbol}

\newcommand\va{\mvector{a}}

\newcommand\vd{\mvector{d}}

\newcommand\vf{\mvector{f}}

\newcommand\vp{\mvector{p}}
\newcommand\vq{\mvector{q}}

\newcommand\vs{\mvector{s}}

\newcommand\vu{\mvector{u}}
\newcommand\vv{\mvector{v}}
\newcommand\vw{\mvector{w}}
\newcommand\vx{\mvector{x}}
\newcommand\vy{\mvector{y}}

\newcommand\vA{\mvector{A}}

\newcommand\vE{\mvector{E}}
\newcommand\vF{\mvector{F}}

\newcommand\vI{\mvector{I}}

\newcommand\vM{\mvector{M}}

\newcommand\vX{\mvector{X}}
\newcommand\vY{\mvector{Y}}

\newcommand\veta{\mvector{\eta}}
\newcommand\vxi{\mvector{\xi}}

\newcommand\vlambda{\mvector{\lambda}}
\newcommand\vLambda{\mvector{\Lambda}}

\newcommand\vGamma{\mvector{\Gamma}}

\newcommand\vomega{\mvector{\omega}}

\newcommand\vvarphi{\mvector{\varphi}}

\newcommand\vXi{\mvector{\Xi}}
\newcommand\vzero{\mvector{0}}

\newcommand\field{\mathbb}

\newcommand\R{\field{R}}

\newcommand\C{\field{C}}
\newcommand\Z{\field{Z}}

\newcommand\N{\field{N}}

\newcommand\T{\field{T}}
\newcommand\bbP{\mathbb{P}}

\newcommand\tr{\operatorname{Tr}}
\newcommand\res{\operatorname{res}}

 \newcommand\spectr{\operatorname{spectr}}

\newcommand\grad{\operatorname{grad}}

\newcommand\card{\operatorname{card}}

\newcommand\rmd{\mathrm{d}}

\newcommand\CP{\ensuremath{\C\bbP}}
\newcommand\rmi{\mathrm{i}\mspace{1mu}}

\newcommand\Dt{\frac{\mathrm{d}\phantom{t} }{\mathrm{d}\mspace{1mu} t}}

\newcommand\pder[2]{\dfrac{\partial #1 }{\partial #2}}
\newcommand\abs[1]{\lvert #1 \rvert}
\newcommand\norm[1]{\lVert #1 \rVert}

\newcommand\pairing[2]{\langle {#1}, {#2}\rangle}
\newcommand\bfi[1]{{\bfseries\itshape{#1}}}

\newcommand\mtext[1]{\quad\text{#1}\quad}

\newcommand\defset[2]{\left\{{#1}\;\vert \;\; {#2} \,\right\}}
\newcommand\deftuple[2]{\left({#1}\;\vert \;\; {#2} \,\right)}

\begin{document}
\thispagestyle{empty}
\vspace*{1em}
\begin{center}
  \Large\textsc{ Differential Galois theory and Integrability  }
\end{center}
\vspace*{0.5em}
\begin{center}
\large  Andrzej J.~Maciejewski$^1$ and Maria Przybylska$^2$
\end{center}
\vspace{1.0em}
\hspace*{2em}\begin{minipage}{0.8\textwidth}
\small
$^1$J.~Kepler Institute of Astronomy,
  University of Zielona G\'ora,
  Licealna 9, \\\quad PL-65--417 Zielona G\'ora, Poland
  (e-mail: maciejka@astro.ia.uz.zgora.pl)\\[0.5em]
$^2$Toru\'n Centre for Astronomy,
   N.~Copernicus University, 
  Gagarina 11, \\\quad PL-87--100 Toru\'n, Poland,
  (e-mail:Maria.Przybylska@astri.uni.torun.pl)
\end{minipage}
\vspace{1.5em}

\begin{abstract}
  This paper is an overview of our works which are related to investigations of  the
  integrability of natural Hamiltonian systems with homogeneous
  potentials and Newton's equations with homogeneous velocity
  independent forces. The two types of integrability obstructions for
  these systems are presented. The first, local ones, are related to
  the analysis of the differential Galois group of variational
  equations along a non-equilibrium particular solution. The second,
  global ones, are obtained from the simultaneous analysis of
  variational equations related to all particular solutions belonging
  to a certain class. The marriage of these two types of the
  integrability obstructions enables to realise the classification
  programme of all integrable homogeneous systems. The main steps of
  the integrability analysis for systems with two and more degrees of
  freedom as well as new integrable systems are shown.
\end{abstract}

keywords:integrability; differential Galois theory; Hamiltonian systems.

\section{Introduction}
It is hard to imagine yourself a world without integrable
models. Teaching in such a world would be rather frustrating.  If a
theory has no solvable examples, it is difficult to explain that it is
useful.  Fortunately, in ours we have the harmonic oscillator---the
guinea-pig which serves as pedagogical example for all theories.

Integrable models are exceptional, but we do not neglect them.  Still,
as it was one century ago, finding a new non-trivially integrable
model is a discovery.

How we can find integrable systems? One way, which seems to be the most
natural, is just a search them in the nature. That is, start with
more or less general model, and look for some special cases when the
model is integrable. The other way is to construct integrable
systems.  It appears that the first approach is much more difficult
than the second one. The reason of this is obvious: we know only few
general methods which give strongly enough, and computable necessary
conditions for the integrability.

In this paper we consider only classical dynamical systems which are
described by ordinary differential equations.  There is no a unique
definition of such systems and there is no a unique method for study
of their integrability. Nevertheless, in this paper we will
concentrate mainly on applications of only one quite new theory.  It
was developed by Baider, Churchill, Morales, Ramis, Rod, Sim\'o and
Singer in the end of the XX century, see
\cite{Audin:01::c,Churchill:96::b,Morales:99::c} and references
therein. In the context of Hamiltonian systems it is called the
Morales-Ramis theory. It arose from very long searching for relations
between the branching of solutions of differential equations
considered as functions of complex time, and the integrability.  In
the context of Hamiltonian systems these relations were found by
S.~L.~Ziglin~\cite{Ziglin:82::b,Ziglin:83::b}. His elegant theory
expresses necessary conditions for the integrability in terms of
properties of the monodromy group of variational equations along a
particular solution.  The Morales-Ramis theory, in some sense, is an
algebraic version of the Ziglin theory. It formulates the necessary
conditions for the integrability in terms of the differential Galois
group of the variational equations.

During one and half decade the Morales-Ramis theory was applied
successfully for study the integrability of numerous systems, see an
overview paper~\cite{Morales:09::}.  Let us mention two biggest
successes of this theory.  It was applied to prove the
non-integrability of the planar three body problem
\cite{Tsygvintsev:01::b,Tsygvintsev:01::a,Boucher:03::,Tsygvintsev:03::},
and to prove the non-integrability of the Hill lunar
problem~\cite{Morales:05::a}.  Moreover, as it is well known, the
first proof of the fact that the problem of the heavy top is
integrable only in the classical cases was done by~S.~L.~Ziglin
in~\cite{Ziglin:83::b} and it is based on his theory. An alternative
proof based on differential Galois approach is given
in~\cite{Maciejewski:05::a}.

The above examples show the real power of the differential Galois
methods in a study of the integrability.  During last ten years we
applied them to study several Hamiltonian and non-Hamiltonian systems
which appear in physics and astronomy, see, e.g.,
\cite{mp:08::d,Maciejewski:06::,MR2123446,MR2191372,Maciejewski:04::,Maciejewski:04::d,Maciejewski:04::a,Maciejewski:03::e,Maciejewski:03::a,Maciejewski:02::b,Maciejewski:02::a}.
Always a hidden dream of those investigations was a strong will to
find an unexpectedly integrable system. However, for a long time,
neither we, nor other authors succeeded with this respect.  All those
successful investigations gave negative results: the investigated
systems are not integrable except already known integrable cases.

In this paper our aim is to give an overview of our works concerning
the integrability of Hamiltonian systems with homogeneous potentials
\cite{mp:08::e,mp:09::b,mp:07::a,Maciejewski:05::b,Maciejewski:04::g},
and homogeneous Newton equations \cite{mp:08::a}.  Our motivation for
those works was an optimistic plan to find the necessary and
sufficient conditions for the integrability. In other words, our dream
was to find all possible integrable polynomial potentials and forces.

Amazingly enough this plan was not only a dream---we found its quite
successful realisation.  Our investigations differ in many respects
from typical applications of the Morales-Ramis theory. In a `typical'
application of this theory there are two difficult problems: the first
one is to find a particular solution, and the second is to determine
the differential Galois group of the variational equations along this
solution.  In this paper we investigate `global' multi-parameter
problems.  We know \emph{a priori} a class of particular solutions and
we know also the differential Galois group for a given values of the
parameters.  The goal in our problem is to prove that the system is
not integrable for all but finite number of parameters' values.  To
achieve the desired result we developed a method which allows to
deduce new obstructions for the integrability that come from a global
analysis of all possible particular solutions of a given class.
 
Plan of this paper is the following. In the next section we overview
various notions of the integrability of ordinary differential
equations. In Section~\ref{sec:gen} we explain how the differential
Galois theory can be used for a study of their integrability.  The
Morales-Ramis theory, as well as the Ziglin theory are dedicated for
Hamiltonian systems.  In our presentation we show that, in fact, we
can use the differential Galois methods in a general context.  One can
find necessary conditions for the integrability defined adequately to
the geometry of the considered equations and express these conditions in
terms of properties of the differential Galois group of variational
equations.  The next two sections~\ref{sec:morahom} and \ref{sec:glob}
are devoted the integrability analysis of the class of homogeneous
potentials. The homogeneity of the potential guarantees the existence
of non-equilibrium particular solutions generated by so-called Darboux
points. In Section~\ref{sec:morahom} the necessary integrability
conditions obtained from an application of the Morales-Ramis theory to
a particular solution given by a Darboux point are formulated. But
these conditions are too weak for the ambitious classification
programme of all integrable potentials. In Section~\ref{sec:glob} the
additional integrability conditions obtained from the simultaneous
analysis of all Darboux points are formulated. In Section~\ref{sec:newtony} it was
shown that various parts of analysis made for homogeneous potentials
can be adapted for systems of Newton equations with homogeneous
velocity-independent forces. In final Section~\ref{sec:open} open
problems and perspectives of the classification program of integrable
systems are discussed.
\section{Integrability}
\label{sec:int}
In this section we overview various notions of the integrability of
systems of ordinary differential equations.

Let us consider a system of ordinary differential equations
\begin{equation}
  \label{eq:ds}
  \Dt\vx=\vv(\vx), \qquad \vx=(x^{1},\ldots,x^{m})    \in\R^{m}, 
\end{equation}
with smooth right-hand sides $\vv(x)=(v^{1}(\vx), \ldots,
v^{m}(\vx))$.  As it was observed a long time ago, the knowledge of
first integrals or other invariant quantities helps to find explicit
solutions of this system.  Let us recall that a smooth function $F$ is
a first integral of system \eqref{eq:ds} iff it is constant along its
solutions.  Thus, a constant value level of $F$ is a set invariant
with respect to the flow of ~\eqref{eq:ds}.  Hence, roughly speaking,
knowing a first integral we can reduce the dimension of the system by
one. Thus, if we know $m-1$ first integrals $F_{1},\ldots, F_{m-1}$,
which are functionally independent in a certain domain $U\subset
\R^{m}$, then we can transform system~\eqref{eq:ds} into the following
one
\begin{equation}
  \label{eq:dsy}
  \Dt \vy =\vw(\vy) , \qquad \vw(\vy)=(0,\ldots,0, w^{m}(\vy)).
\end{equation}
Solutions of this equation can be find easily
\begin{equation}
  \label{eq:sol}
  y^{i}(t):= y^{i}_{0} \mtext{for} i=1, \ldots, m-1,  
\end{equation}
and $ y^{m}(t)$ is given by
\begin{equation}
  \label{eq:if}
  \int_{y^{m}_{0}}^{y^{m}(t)} \frac{\rmd s}{w^{m}(y^{1}_{0},\ldots,
    y^{m-1}_{0},s)}=t. 
\end{equation}
In effect, we reduce the problem to calculation just one integral and
inversion of a function. In the prescribed situation, we say that the
considered system is \emph{integrable by quadratures}.  This notion
plays a fundamental role. In fact, as we will see later, the other
definitions of  integrability give necessary conditions for the
integrability by quadratures.

A first integral of a given system is just an example of simplest
tensor quantity which is invariant with respect to the flow of the
system.  A smooth tensor field $T(\vx)$ is invariant with respect to the
flow of the system~\eqref{eq:ds} iff
\begin{equation}
  \label{eq:invt}
  L_{\vv}(T)=0,
\end{equation}
where $ L_{\vv}$ denotes the Lie derivative along vector field $\vv$.
The existence of a certain number of tensor invariants can guarantee
that the integration of the system reduces to quadratures. Below we
give three examples of results of this type.  The first of them, due
to S.~Lie, tells that system admitting $m$ linearly independent and
commuting symmetries, i.e., invariant vector fields, is integrable by
quadratures.
\begin{theorem}[S.~Lie]
  Assume that system~\eqref{eq:ds} admits $m$ independent and
  commuting symmetries $\vu_{1}=\vv$, $\vu_{2}, \ldots, \vu_{m}$. Then
  the system is integrable by quadratures.
\end{theorem}
A differential $m$-form $\omega$ in $\R^{m}$ is given by
\begin{equation}
  \label{eq:o}
  \omega= h(\vx)\,\rmd x^{1}\wedge \cdots \wedge \rmd x^{m}. 
\end{equation}
It is invariant with respect to system~\eqref{eq:ds} iff
\[
L_{\vv}(\omega)=\left(\sum_{i=1}^m\dfrac{\partial (h v^{i})}{\partial
    x^{i}}\right)\mathrm{d}x^{1}\wedge\cdots\wedge\mathrm{d}x^{m}=0.
\]
In the classical literature an invariant $m$-form is called the Jacobi
last multiplier.

\begin{theorem}[L.~Euler, C.~G.~J.~Jacobi]
  Assume that system~\eqref{eq:ds} admits $m-2$ functionally
  independent first integrals and invariant differential
  $m$-form. Then the system is integrable by quadratures.
  \label{thm:nic}
\end{theorem}
Proofs and detailed discussion of the above two theorems can be find,
e.g., in \cite{Kozlov:96::,Trofimov:95::}.

As we will see later, the following theorem, due to
O.~I.~Bogoyavlensky \cite{MR97k:58147,Bogoyavlenskij:96::a}, is very
important in the context of applications of differential Galois method
to the integrability of non-Hamiltonian systems.
\begin{theorem}[O.~I.~Bogoyavlensky]
  \label{thm:bogo}
  Assume that system~\eqref{eq:ds} admits $1\leq k<m$ functionally
  independent first integrals $F_{1},\ldots,F_{k}$, and $m-k$ linearly
  independent and commuting symmetries $\vu_{1}=\vv$, $\vu_{2},
  \ldots, \vu_{m-k}$, such that
  \begin{equation}
    \label{eq:fi1}
    L_{\vu_{j}}(F_{i})=\vu_{j} [ F_{i} ]=0, \mtext{for}  1\leq i\leq k; \quad 1\leq j\leq m-k.
  \end{equation}
  Then the system is integrable by quadratures.
\end{theorem}
Assumptions of the above theorems can serve as a source of definitions
for specific types of the integrability. Thus, for example, we have
the following definition which is frequently used in non-holonomic
mechanics.
\begin{definition}
  \label{def:jac}
  We say that system~\eqref{eq:ds} is integrable in the Jacobi sense
  iff it admits $m-2$ independent first integrals and invariant
  differential $m$-form.
\end{definition}
In a similar way, we can take the assumptions of
Theorem~\ref{thm:bogo} as a basis for the definition of
$B$-integrability.
\begin{definition}
  We say that system~\eqref{eq:ds} is $B$-integrable iff it admits
  $1\leq k\leq m$ functionally independent first integrals $F_1,
  \ldots, F_k$, and $(m-k)$ linearly independent and commuting
  symmetries $\vu_{1}=\vv, \ldots, \vu_{m-k}$, such that
  $\vu_j[F_i]=0$ for $1\leq i\leq k$, $1\leq j \leq m-k$.
\end{definition}
The above definition arises from a careful analysis of the concept of
the integrability of Hamiltonian systems in the Liouville sense (see
below).

To describe shortly the most characteristic features of $B$ integrable
systems  let us consider such a system, and let us assume that it admits
functionally independent first integrals $F_{1}, \ldots, F_{k}$. With
these integrals we associate the momentum map
\begin{equation}
  \label{eq:m}
  \R^{m}\ni \vx \longmapsto \vF(\vx):=(F_{1}(\vx), \ldots,F_{k}(\vx))\in \R^{k}.
\end{equation}
Let us consider a common level of the first integrals
\begin{equation}
  \label{eq:cl}
  M_{\vf}:=\vF^{-1}(\vf)=\defset{\vx\in\R^{m}}{\vF(\vx)=\vf},
\end{equation}
where $\vf\in\R^{k}$.  If first integrals are independent on
$M_{\vf}$, then $M_{\vf}$ is a smooth manifold. But, even if $M_{\vf}$
is connected and compact, its topology can be very
complicated. However, by the assumed $B$-integrability, we know that
there exist independent and commuting vector fields $\vu_{1}, \ldots,
\vu_{m-k}$ tangent to $M_{\vf}$, . Thus, if $M_{f}$ is connected and
compact it is diffeomorphic to $m-k$ dimensional torus $\T^{m-k}$, see
Chapter 10, Lemma~2 in \cite{Arnold:89::}.  In a neighbourhood of
$M_{\vf}$ we can introduce local coordinates $(\vvarphi, \vI)$, where
$\vI\in D\subset \R^{k}$, and $\vvarphi=(\varphi_{1}, \ldots,
\varphi_{m-k})$ are angular coordinates on $\T^{m-k}$. In these
coordinates system~\eqref{eq:ds} reads
\begin{equation}
  \label{eq:aa}
  \Dt\vvarphi=\vomega(\vI), \qquad \Dt \vI =0. 
\end{equation}
Thus, a solution of this system has the form
\begin{equation}
  \label{eq:saa}
  \vI(t):=\vI_{0} \qquad  \vvarphi(t) := \vomega(\vI_{0})t +\vvarphi_{0}.
\end{equation}
From the above considerations it follows that $B$-integrability is
similar to the integrability in the Liouville sense of Hamiltonian
systems. In fact, the $B$-integrability was introduced as a certain
generalisation of the integrability in the Liouville sense.

Let us assume that system~\eqref{eq:ds} is Hamiltonian. Then $m=2n$,
and $\vv=\vX_H$ is a Hamiltonian vector field generated by a smooth
Hamiltonian function $H:\R^{2n}\rightarrow \R$.  Here we consider
$\R^{2n}$ as a linear symplectic space with chosen canonical
coordinates $\vx=(\vq,\vp)$, where $\vq=(q_1,\ldots,q_n)$ and
$\vp=(p_{1},\ldots,p_{n})$. In these coordinates the symplectic form
$\Omega$ is following
\[
\Omega=\sum\limits_{i=1}^n\rmd q_i\wedge\rmd p_i,
\]
and the vector field $\vX_H$ reads
\[
\vX_H=\sum\limits_{i=1}^n \left( \pder{H}{p_i}\pder{\phantom{p}}{q_i}
  - \pder{H}{q_i}\pder{\phantom{p}}{p_i} \right).
\]
Let us recall the definition of the well known notion.
\begin{definition}
  We say that Hamiltonian vector field $\vX_H$ is integrable in the
  Liouville sense iff it admits $n$ functionally independent and
  commuting smooth first integrals $F_{1}, \ldots, F_{n}$.
\end{definition}
We notice here that a Hamiltonian vector field $X_{H}$ integrable in
the Liouville sense is $B$-integrable. In fact, $X_{H}$ admits $n$
first integrals, and $n=2n-n$ symmetries $\vX_{F_{1}}, \ldots,
\vX_{F_{n}} $, which satisfy
\begin{equation}
  \vX_{F_{i}}[F_{j}]=\{F_{i},F_{j}\}=0 \mtext{for}1\leq i,j\leq n,
\end{equation}
where $\{\cdot,\cdot\}$ denotes the Poisson bracket.

Obviously, a $B$-integrable Hamiltonian system can be non-integrable
in the Liouville sense.

\section{General theory}
\label{sec:gen}
In this section we show how we can use the differential Galois theory
to find necessary conditions for the integrability of ordinary
differential equations.  To deduce such conditions we have to make
several assumptions. The most general one is that the considered
system as well as the considered first integrals, or other invariants,
have `good' analytical properties.  Moreover, the theory requires that
the `scalars' form an algebraically closed field. Thus, we assume that
this field is just the field of complex numbers $\C$. In effect we
work with complex functions, complex vector fields, and so on.  The
above mentioned `good' analytical properties mean that the considered
tensors are holomorphic at points where they are defined.

Let us consider a complex holomorphic system of ordinary differential
equations
\begin{equation}
  \label{eq:hds}
  \Dt \vx = \vv(\vx), \qquad \vx\in U \subset \C^{m}, \quad t\in\C, 
\end{equation}
where $U$ is an open and connected subset of $\C^{m}$.  The basic
assumption for further considerations is that we know a particular
non-equilibrium solution $\vvarphi(t)$ of this system.  Usually it is
not a single-valued function of the complex time $t$. Thus, we
associate with $\vvarphi(t)$ a Riemann surface $\Gamma$ with $t$ as a
local coordinate.

The variational equations along $\Gamma$ have the form
\begin{equation}
  \label{eq:vds}
  \Dt\vxi = \vA(t) \vxi, \qquad  \vA(t)=\frac{\partial
    \vv}{\partial \vx}(\vvarphi(t) ) ,  \qquad \vxi \in T_\Gamma U.
\end{equation}
The entries of matrix $\vA(t)$ in the above equation are elements of
field $K:=\mathscr{M}(\Gamma)$ of functions meromorphic on $\Gamma$.
This field with the differentiation with respect to $t$ as a
derivation is a differential field. Only constant functions from $K$
have vanishing derivative, so the sub-field of constants of $K$ is
$\C$.

It is obvious that solutions of~\eqref{eq:vds} are not necessarily
elements of $K^m$. The fundamental theorem of the differential Galois
theory guarantees that there exists a differential field $L\supset K$
such that $m$ linearly independent (over $\C$) solutions
of~\eqref{eq:vds} are contained in $L^m$. The smallest differential
extension $L\supset K$ with this property is called the Picard-Vessiot
extension of $K$.

A group $\mathscr{G}$ of differential automorphisms of $L$ which do
not change $K$ is called the differential Galois group of
equation~\eqref{eq:vds}.  It can be shown that $\mathscr{G}$ is a
linear algebraic group.  Thus, in particular, it is a union of a
finite number of disjoint connected components. One of them,
containing the identity, is called the identity component and is
denoted by $\mathscr{G}^{\circ}$.

Let $\vxi=(\xi^{1}, \ldots,\xi^{m})^T\in L^m$ be a solution of
equation~\eqref{eq:vds}, and $g$ an element of its differential Galois
group $\scG$.  Then, $g(\vxi):= (g(\xi^{1}), \ldots,g(\xi^{m}))^T$ is
also its solution. In fact, by definition $g$ commutes with the time
differentiation, so we have
\begin{equation*}
  \Dt g(\vxi)= g \left( \Dt \vx \right) = g\left(\vA(t)\vxi\right)=g(\vA(t))g(\vxi)=\vA(t)g(\vxi),
\end{equation*}
as $g$ does not change elements of $K$. Thus, if $\vXi\in
\mathrm{GL}(n,L)$ is a fundamental matrix of~\eqref{eq:vds}, i.e., its
columns are linearly independent solutions of~\eqref{eq:vds}, then $
g(\vXi)=\vXi \vM_g $, where $\vM_g\in \mathrm{GL}(n,\C)$. In other
words, the differential Galois group $\scG$ can be considered as an
algebraic subgroup of $\mathrm{GL}(m,\C)$.

Now, we explain how the existence of first integrals of system
\eqref{eq:hds} manifests itself in the properties of the differential
Galois group of variational equations.  At first, we introduce some
definitions.  Let us consider a holomorphic function $F$ defined in a
certain connected neighbourhood of solution $\vvarphi(t)$. In this
neighbourhood we have the expansion
\begin{equation}
  F(\vvarphi(t) +\vxi)= F_l(\vxi) + O(\norm{\vxi}^{l+1}),  \qquad F_l\neq 0.
\end{equation}
Then, the leading term $f$ of $F$ is the lowest order term of the
above expansion i.e., $f(\vxi):=F_l(\vxi)$.  Note that $f(\vxi)$ is a
homogeneous polynomial of variables $\vxi=(\xi^{1},\ldots, \xi^{m})$
of degree $l$.  If $F$ is a meromorphic function, then it can be
written as $F=P/Q$ for certain holomorphic functions $P$ and $Q$.
Then the leading term $f$ of $F$ is defined as $f=p/q$, where $p$ and
$q$ are leading terms of $P$ and $Q$, respectively. In this case
$f(\vxi)$ is a homogeneous rational function of $\vxi$.

It is easy to prove that if $F$ is a meromorphic (holomorphic) first
integral of equation~\eqref{eq:hds}, then its leading term $f$ is a
rational (polynomial) first integral of variational
equation~\eqref{eq:vds}. If system ~\eqref{eq:hds} has $k\geq 2$
functionally independent meromorphic first integrals $F_1, \ldots,
F_k$, then their leading terms can be functionally dependent. However,
by the Ziglin Lemma \cite{Ziglin:82::b,Audin:01::c,Churchill:96::b},
we can find $k$ polynomials $G_1, \ldots, G_k\in\C[z_1,\ldots, z_k]$
such that leading terms of $G_i(F_1,\ldots, F_k)$, for $1\leq i \leq
k$ are functionally independent.

Additionally, if $\mathscr{G}\subset \mathrm{GL}(m,\C)$ is the
differential Galois group of~\eqref{eq:vds}, and $f$ is its rational
first integral, then $f(g(\vxi))= f(\vxi)$ for every
$g\in\mathscr{G}$, see \cite{Audin:01::c,Morales:99::c}, i.e., $f$ is
a rational invariant of group $\mathscr{G}$. Thus we have a
correspondence between the first integrals of the
system~\eqref{eq:hds} and invariants of $\mathscr{G}$.
\begin{lemma}
  If equation~\eqref{eq:hds} has $k$ functionally independent first
  integrals which are meromorphic in a connected neighbourhood of a
  non-equilibrium solution $\varphi(t)$, then the differential Galois
  group $\mathscr{G}$ of the variational equations along $\vvarphi(t)$
  has $k$ functionally independent rational invariants.
  \label{lem:ratinv}
\end{lemma}
As it was mentioned above, a differential Galois group is a linear
algebraic group, thus, in particular, it is a Lie group, and one can
consider its Lie algebra. This Lie algebra reflects only the
properties of the identity component of the group.  It is easy to show
that if a Lie group has an invariant, then also its Lie algebra has an
integral. Let us explain what the last sentence means.  Let
$\efg\subset \mathrm{gl}(m,\C)$ denote the Lie algebra of
$\mathscr{G}$. Then, an element $\vY\in \efg\subset \mathrm{gl}(m,\C)$
can be considered as a linear vector field: $\vx\mapsto \vY(\vx):=
\vY\cdot \vx$, for $\vx\in\C^m$.  We say that $f\in\C(x_{1}, \dots,
x_m)$ is an integral of $\efg$, iff $\vY(f)(\vx)=\rmd f(\vx)\cdot
\vY(\vx)=0$, for all $\vY\in \efg$.
\begin{proposition}
  \label{prop:lieint}
  If $f_1, \ldots, f_k\in\C(x_1, \ldots,x_m)$ are algebraically
  independent invariants of an algebraic group $\mathscr{G}\subset
  \mathrm{GL}(m,\C)$, then they are algebraically independent first
  integrals of the Lie algebra $\efg$ of $\mathscr{G}$.
\end{proposition}

The above facts are the starting points for applications of
differential Galois methods to a study of the integrability.

If the considered system is Hamiltonian, then we have additional
constrains.  First of all, the differential Galois group of
variational equations is a subgroup of the symplectic group
$\mathrm{Sp}(2n,\C)$. Secondly, commutation of first integrals imposed
by the Liouville integrability implies commutation of first integrals
of variational equations.  The following lemma plays the crucial role
and this is why it was called The Key Lemma, see Lemma~III.3.7 on page
72 in~\cite{Audin:01::c}.
\begin{lemma}
  \label{lem:key}
  Assume that Lie algebra $\efg\subset \mathrm{sp}(2n,\C)$ admits $n$
  functionally independent and commuting first integrals. Then $\efg$
  is Abelian.
\end{lemma}
Hence, if $\efg$ in the above lemma is the Lie algebra of a Lie group
$\mathscr{G}$, then the identity component $\mathscr{G}^\circ$ of
$\mathscr{G}$ is Abelian.

Using all these facts Morales and Ramis proved the following theorem
\cite{Morales:99::c,Morales:01::b1}.
\begin{theorem}[Morales-Ramis]
  \label{thm:basicG}
  Assume that a Hamiltonian system is meromorphically integrable in
  the Liouville sense in a connected neighbourhood of a phase curve
  $\Gamma$. Then the identity component of the differential Galois
  group of the variational equations along $\Gamma$ is Abelian.
\end{theorem}

If the considered system is not Hamiltonian, then there is no a
commonly accepted definition of the integrability.  However, if we
restrict yourself to $B$-integrability, then we have a beautiful
generalization of Theorem~\ref{thm:basicG}.  Namely, with
system~\eqref{eq:hds} we consider also its cotangent lift, i.e., a
Hamiltonian system defined in $\C^{2m}$ by Hamiltonian function
\begin{equation}
  \label{eq:clh}
  H=\sum_{i=1}^{m} y_{i}v^{i}(\vx)
\end{equation}
where $(\vx,\vy)=(x^{1},\ldots,x^{m},y_{1}, \ldots, y_{m})$ are
canonical coordinates in $\C^{2m}$.  Thus, the Hamiltonian equations
have the form
\begin{equation}
  \label{eq:hh}
  \Dt x^{i}=\pder{H}{y_{i}}=v^{i}(\vx), \qquad \Dt
  y_{i}=-\pder{H}{x^{i}}=-\sum_{j=1}^{m} y_{j}\pder{v^{j}}{x^{i}}(\vx),
  \quad 1\leq i \leq m. 
\end{equation}
Let us assume that system~\eqref{eq:hds} is $B$-integrable with $k$
first integrals $F_{1}, \ldots, F_{k}$, and $m-k$ symmetries $\vu_{1},
\ldots, \vu_{m-k}$.  Then, we claim that Hamiltonian
system~\eqref{eq:hh} is integrable in the Liouville sense.  Let us
define the following functions
\begin{equation}
  \label{eq:ff}
  F_{k+j}(\vx,\vy): =\pairing{\vy}{\vu_{j}(\vx)}:= 
  \sum_{i=1}^{m}y_{i}u_{j}^{i}(\vx) \mtext{for} 1\leq j\leq
  m-k. 
\end{equation}
We have
\begin{equation}
  \label{eq:tfj}
  \{F_{j+k},H\}=\sum_{i=1}^{m}\left( \pder{F_{j+k}}{x^{i}}
    \pder{H}{y_{i}}-  \pder{F_{j+k}}{y_{i}}
    \pder{H}{x^{i}}\right)=\pairing{\vy}{[\vu_{j},\vv]}. 
\end{equation}
But, $[\vu_{j},\vv]=0$ for $1\leq j\leq m-k$, by assumption, and so
$F_{j+k}$ are first integrals. We show that first integrals $F_{1},
\ldots, F_{m}$ pairwise commute. Obviously, we have
$\{F_{i},F_{j}\}=0$, for $1\leq i,j\leq k$. Moreover, we have
\begin{equation}
  \label{eq:cc}
  \{F_{j+k},F_{i+k}\}=\pairing{\vy}{[\vu_{j},\vu_{i}]}=0 \mtext{for} 1\leq
  i,j\leq m-k, 
\end{equation}
as, by assumption, $[\vu_{j},\vu_{i}]=0$. Finally, we have also
\begin{equation}
  \label{eq:ccc}
  \{F_{i},F_{j+k}\}= \vu_{j}[ F_{i} ]=0, \mtext{for} 1\leq i\leq k, \mtext
  1\leq j\leq m-k,
\end{equation}
because, by assumption, $F_1, \ldots, F_k$ are common first integrals
of $\vu_1, \ldots, \vu_{m-k}$.  Thus, we proved our claim.

Now, let $\vvarphi(t)=(\varphi^1(t),\ldots, \varphi^m(t))$ be a
particular solution of \eqref{eq:hds}. Then,
\[
t\longmapsto \widetilde\vvarphi(t):= (\vvarphi(t),\vzero)\in\C^{2m},
\]
is a particular solution of the Hamilton equation \eqref{eq:hh}.  The
variational equations for this solution have the following form
\begin{equation}
  \label{eq:vhh}
  \Dt \vxi=\vA(t)\vxi, \qquad \Dt \veta=-\vA(t)^T\veta, \qquad  \vA(t)=\frac{\partial
    \vv }{\partial \vx}(\vvarphi(t) ). 
\end{equation}
The first of the above equations is just the variational
equation~\eqref{eq:vds}, and the second one is its adjoint.  Thus, if
$\vXi$ is a fundamental matrix of the first equation in
\eqref{eq:vhh}, then $\vX:=(\vXi^{-1})^T$ is a fundamental matrix of
the second equation in \eqref{eq:vhh}. In effect, the differential
Galois group of system \eqref{eq:vhh} coincides with the differential
Galois group of its first equation, i.e., with the differential Galois
group of the original variational equtions \eqref{eq:vds}.  Using the
above facts Ayoul and Zung proved in \cite{Ayoul:09::} the following
theorem.
\begin{theorem}[Ayoul-Zhung]
  \label{thm:zunG}
  Assume that system~\eqref{eq:hds} is meromorphically $B$-integrable
  in a connected neighbourhood of a phase curve $\Gamma$. Then the
  identity component of the differential Galois group of the
  variational equations along $\Gamma$ is Abelian.
\end{theorem}
Let us underline the importance of this theorem. All results which
were obtained on the basis of Theorem~\ref{thm:basicG} and stating
that a given Hamilton system is non-integrable in the Liouville sense
are, in fact, much stronger -- the considered systems are not
$B$-integrable.
Already in the book \cite{Morales:99::c} a very natural extension of
described approach was presented. Except the variational
equations~\eqref{eq:vds} along the phase curve $\Gamma$ corresponding
to the particular solution $\vvarphi(t)$, we can consider also the
higher order variational equations.  To derive them we consider
system~\eqref{eq:hds} in a neighbourhood of $\Gamma$, where we can
write the following expansion
\[
\vx=\vvarphi(t)+\varepsilon\vxi^{(1)}+\varepsilon^2\vxi^{(2)}+\cdots+
\varepsilon^k\vxi^{(k)}+\cdots,
\]
where $\varepsilon $ is a formal small parameter. Inserting the above
expansion into equation~\eqref{eq:hds} and collecting terms of the
same order with respect to $\varepsilon$, we obtain the following
chain of equations
\begin{equation}
  \label{eq:vek}
  \Dt\vxi^{(k)} = A(t)\vxi^{(k)} + \vf_k(\vxi^{(1)},
  \ldots, \vxi^{(k-1)}), \qquad k=1,2, \ldots,
\end{equation}
where $\vf_1\equiv 0$. For a fixed $k$, this is a system of $k$-th
order variational equations, and we denote it by $\mathrm{VE}_{k}$.
It is a linear and, for $k>1$, non-homogeneous system of
equations. Nevertheless, there exists an appropriate framework
allowing to define the differential Galois group $\scG_{k}$ of
$\mathrm{VE}_{k}$ for any $k$.  Obviously, $\mathrm{VE}_{1}$ coincides
with~\eqref{eq:vds}, so $\scG_{1}$ coincides with $\scG$. A detailed
exposition and proofs the reader will find in~\cite{Morales:07::},
where among other things the following generalisation of
Theorem~\ref{thm:basicG} is given.
\begin{theorem}[Morales-Ramis-Sim\'o]
  \label{thm:basicGk}
  Assume that a Hamiltonian system is meromorphically integrable in
  the Liouville sense in a connected neighbourhood of a phase curve
  $\Gamma$. Then the identity component of the differential Galois
  group $\scG_{k}$ of $k$-th variational equations $\mathrm{VE}_{k}$
  along $\Gamma$ is Abelian, for all $k\in\N$.
\end{theorem}
Hence, Theorem~\ref{thm:basicG} gives only the first order
obstructions for the integrability.  If $\scG^{\circ}=\scG_{1}^{\circ}
$ is Abelian, then, having only Theorem~\ref{thm:basicG}, we cannot be
sure whether the system is integrable or not. But knowing the above
theorem we can continue our investigations and check if the
$\scG_{2}^{\circ}$ is Abelian. If it is not, the system is not
integrable, otherwise we have to check if $\scG_{3}^{\circ}$ is
Abelian. This process we continue up to such $k$ that $\scG_i^{\circ}$
is Abelian for $i<k$, and $\scG_k^{\circ}$ is not Abelian. If we are
able to find such $k$, then the system is not integrable.

Here it is worth to mention that it is very hard to determine the
differential Galois groups $\scG_{k}$ with $k>1$, or even to decide
whether $\scG_{k}^{\circ}$ is Abelian or not. This is why we have only
a few applications of Theorem~\ref{thm:basicGk},
see~\cite{Morales:07::,Morales:09::}.  However, all successful
applications of this theorem show its real power.

We can derive higher order variational equations for an arbitrary
system~\eqref{eq:hds}.  Thus, we can ask if we have a generalisation
of Theorem~\ref{thm:zunG}, similar to that described above for
Theorem~\ref{thm:basicGk}.  In fact, we have such generalisation.
\begin{theorem}[Ayoul-Zhung]
  \label{thm:zunGk}
  Assume that system~\eqref{eq:hds} is meromorphically $B$-integrable
  in a connected neighbourhood of a phase curve $\Gamma$. Then the
  identity component $\scG_{k}^{\circ}$ of the differential Galois
  group $\scG_{k}$ of the $k$-th variational equations along $\Gamma$
  is Abelian, for all $k\in\N$.
\end{theorem}
\section{Integrability of homogeneous potentials -- Morales-Ramis
  theorem}
\label{sec:morahom}
Let us consider Hamiltonian systems with $n$ degrees of freedom given
by a natural Hamiltonian function
\begin{equation}
  \label{eq:ham}
  H = \frac{1}{2}\sum_{i=1}^np_i^2 +V(\vq)\,
\end{equation}
where $\vq=(q_1,\ldots,q_n)$ and $\vp=(p_1,\ldots,p_n)$ are canonical
coordinates and momenta, $V$ is a homogeneous function of degree
$k\in\Z^{\star}:=\Z\setminus\{0\}$.  We assume just from the beginning that the
considered system is complex, i.e., the phase space is $\C^{2n}$.  The
Hamilton equations have the canonical form
\begin{equation}
  \label{eq:heqs}
  \Dt \vq = \vp, \qquad \Dt \vp = -V'(\vq),
\end{equation}
where $V'(\vq):=\grad V(\vq)$. Morever, we assume also that the time
$t$ is a complex variable.

We say that a potential $V$ is integrable iff the Hamiltonian
system~\eqref{eq:heqs} is integrable. 

One of the most beautiful applications of the Morales-Ramis
Theorem~\ref{thm:basicG} concerns Hamiltonian systems of the
prescribed above form.  The basic assumption in this application is
that there exists a non-zero vector $\vd\in\C^{n}$ such that
\begin{equation}
  \label{eq:dar}
  V'(\vd)=\vd,
\end{equation}
It is is called a proper Darboux point of potential $V$.
It defines a two dimensional plane in the phase spacs $\C^{2n}$, given
by
\begin{equation}
  \Pi(\vd):= \defset{(\vq,\vp)\in\C^{2n}}{ \vq =\varphi\vd, \  \vp=\psi\vd,
    \quad (\varphi,\psi)\in\C^2 }.
\end{equation}
This plane is invariant with respect to the system~\eqref{eq:heqs}.
Equations~\eqref{eq:heqs} restricted to $\Pi(\vd)$ have the form of
one degree of freedom Hamilton's equations
\begin{equation}
  \label{eq:one}
  \Dt \varphi = \psi, \qquad \Dt \psi = -\varphi^{k-1},
\end{equation}
with the following phase curves
\begin{equation}
  \Gamma_{k,\varepsilon} :=\defset{(\varphi,\psi)\in\C^2}{ \frac{1}{2}\psi^2 +\frac{1}{k}\varphi^k=
\varepsilon}\subset\C^2, \qquad \varepsilon \in\C.
\end{equation}
In this way, a solution $(\varphi,\psi)=(\varphi(t),\psi(t))$
of~\eqref{eq:one} gives rise a solution
$(\vq(t),\vp(t)):=(\varphi\vd,\psi\vd)$ of equations~\eqref{eq:heqs}
with the corresponding phase curve
\begin{equation}
  \vGamma_{k,\varepsilon} :=\defset{(\vq,\vp)\in\C^{2n}}{ (\vq,\vp)=(\varphi \vd,\psi\vd),\ (\varphi,\psi)\in  
\Gamma_{k,\varepsilon} }\subset\Pi(\vd).
\end{equation}
Morales and Ramis obtained necessary conditions for the integrability
in the Liouville sense by an analysis of the variational equations
along an arbitrary phase curve $\vGamma_{k,\varepsilon}$ with
$\varepsilon\neq 0$. These variational equations have the form
\begin{equation}
  \label{eq:vart}
  \ddot \vx = -\varphi(t)^{k-2} V''(\vd)\vx,
\end{equation}
where $V''(\vd)$ is the Hessian of $V$ calculated at $\vd$. Let us
assume that $V''(\vd)$ is diagonalisable. Then, without loss of the
generality, we can assume that $V''(\vd)$ is diagonal, and in such a case
system~\eqref{eq:vart} splits into a direct product of second order
equations
\begin{equation}
  \label{eq:varts}
  \ddot x_i = -\lambda_i \varphi(t)^{k-2} x_i, \qquad 1\leq i\leq n,
\end{equation}
where $\lambda_1,\ldots, \lambda_n$ are eigenvalues of $V''(\vd)$.  It
is easy to show that $\vd$ is an eigenvector of $V''(\vd)$ with
eigenvalue $k-1$. We always denote this eigenvalue as $\lambda_n$.

In~\cite{Morales:01::a} J.~J.~Morales-Ruiz and J.~P.~Ramis proved the
following theorem.
\begin{theorem}[Morales-Ramis]
  \label{thm:MoRa}
  Assume that the Hamiltonian system defined by Hamiltonian
  \eqref{eq:ham} with a homogeneous potential $V\in\C(\vq)$ of degree
  $k\in\Z^\star$ satisfies the following conditions:
  \begin{enumerate}
  \item there exists a non-zero $\vd\in\C^n$ such that $V'(\vd)=\vd$,
    and
  \item matrix $V''(\vd)$ is diagonalisable with eigenvalues $\lambda_1,
    \ldots, \lambda_n$;
  \item the system is integrable in the Liouville sense with first
    integrals which are meromorphic in a connected neighbourhood $U$
    of phase curve $\vGamma_{k,\varepsilon}$ with $\varepsilon\neq0$,
    and independent on $U\setminus \vGamma_{k,\varepsilon}$.
  \end{enumerate}
  Then each pair $(k,\lambda_i)$ for $i=1,\ldots,n$ belongs to an item
  of the following list
  \begin{equation}
    \label{eq:tab}
    \begin{array}{crcr}
      \toprule
      \text{case} &  k & \lambda  &\\
      \midrule 
      \vbox to 1.3em{}1. &  \pm 2 &  \text{arbitrary} & \\
\addlinespace[0.5em]
      2.  & k & p + \dfrac{k}{2}p(p-1) & \\
\addlinespace[0.5em]
%\midrule 
      3. & k & \dfrac 1 {2}\left(\dfrac {k-1} {k}+p(p+1)k\right)  & \\
\addlinespace[0.5em]
%\midrule 
      4. & 3 &  -\dfrac 1 {24}+\dfrac 1 {6}\left( 1 +3p\right)^2, & -\dfrac 1
      {24}+\dfrac 3 {32}\left(  1  +4p\right)^2 \\
\addlinespace[0.5em]
      & & -\dfrac 1 {24}+\dfrac 3 {50}\left(  1  +5p\right)^2,  &
      -\dfrac1{24}+\dfrac{3}{50}\left(2 +5p\right)^2 \\
\addlinespace[0.5em]
      5. & 4 & -\dfrac 1 8 +\dfrac{2}{9} \left( 1+ 3p\right)^2 & \\
\addlinespace[0.5em]
      6. & 5 & -\dfrac 9 {40}+\dfrac 5 {18}\left(1+ 3p\right)^2, & -\dfrac 9
      {40}+\dfrac 1 {10}\left(2
        +5p\right)^2\\
\addlinespace[0.5em]
      7. & -3 &\dfrac {25} {24}-\dfrac 1 {6}\left( 1 +3p\right)^2, & \dfrac {25}
      {24}-\dfrac 3 {32}\left(1 +4p\right)^2 \\
\addlinespace[0.5em]
      & & \dfrac {25} {24}-\dfrac 3 {50}\left(1+ 5p\right)^2, & \dfrac {25}
      {24}-\dfrac{3}{50}\left(2+ 5p\right)^2\\
\addlinespace[0.5em]
      8. & -4 & \dfrac 9 8-\dfrac{2}{9}\left( 1+ 3p\right)^2 & \\
\addlinespace[0.5em]
      9. & -5 & \dfrac {49} {40}-\dfrac {5} {18}\left(1+3p\right)^2 ,& \dfrac {49}
      {40}-\dfrac {1} {10}(2 +5p)^2\\
\addlinespace[0.5em]
      \bottomrule
    \end{array}
  \end{equation}
  where $p$ is an integer.
\end{theorem}
Let us remark that the above theorem does not give any obstruction for
the integrability if $k=2$ or $k=-2$.
\begin{remark}
  \label{rem:nd}
  It was explained in \cite{Maciejewski:09::} that the assumption that
  $V''(\vd)$ is diagonalisable is irrelevant. That is, the necessary
  conditions for the integrability are the same: if the potential is
  integrable, then each $\lambda\in\spectr V''(\vd)$ must belong to
  appropriate items of the above list. Additionally, if $V''(\vd)$ is
  not diagonalisable, then new obstacles for the integrability
  appear. Namely, if the Jordan form of $V''(\vd)$ has a block
  \begin{equation*}
    J_3(\lambda):=
    \begin{bmatrix}
      \lambda &1&0\\
      0&\lambda &1\\
      0&0&\lambda
    \end{bmatrix},
  \end{equation*}
  then the system is not integrable. Moreover, if the Jordan form of
  $V''(\vd)$ has a two dimensional block $J_2(\lambda)$, and $\lambda$
  belongs to the second item of table~\eqref{eq:tab}, then the system
  is not integrable. This fact was proved in~\cite{Maciejewski:09::}.
 In other words, for $k\not\in\{-2,0,+2\}$, the presence of a proper Darboux point $\vd$ for
  which  $V''(\vd)$ has a block of dimension greater than two, or
  block of dimension two 
  with corresponding  $\lambda_i$  belonging to the second item of
  table~\eqref{eq:tab},  implies immediately the non-integrability of
  the potential.
\end{remark}
\begin{remark}
  The case of a homogeneous potential of degree $k=0$ needs a special
  treatment. Necessary conditions  for the integrability in this case were found recently
  in~\cite{mp:09::d}.
\end{remark}
We denote by $\mathscr{M}_k$ a subset of rational numbers $\lambda$ specified by
the table in the above theorem for a given $k$, e.g., for $\abs{k}>5$, we have
\begin{equation}
 \mathscr{M}_k=\defset{ p + \dfrac{k}{2}p(p-1)}{ p\in\Z}\cup
\defset{\dfrac 1
{2}\left[\dfrac {k-1} {k}+p(p+1)k\right]}{p\in\Z}.
\end{equation}

The Morales-Ramis Theorem~\ref{thm:MoRa} is a powerful tool. In fact,
the necessary conditions for the integrability of polynomial potentials
are reduced to solving algebraic equations: we have to find a Darboux
point and then to check if the eigenvalues of the Hessian at this
Darboux point are rational numbers which belong to the Morales-Ramis
table.  However, we note that none of these algebraic problems is 
trivial. First of all, even if a Darboux point exists, generally we are
not able to find its coordinates. Moreover, we have much more serious
problems in a parametric case which is the most important in
applications. This is  illustrated by the following
example.

Let us consider the following potential 
\begin{equation}
\label{eq:vac}
 V = \frac{1}{3}a q_1^3 + \frac{1}{2}q_1^2 q_2 + \frac{1}{3}cq_2^3,
\end{equation}
where $a$ and $c$ are in general complex parameters. For generic values of these
parameters $V$ has three Darboux points $\vd_1$, $\vd_2$, and  $\vd_3$.
The non-trivial eigenvalues $\lambda_i = \tr
 V''(\vd_i) -2$ of Hessian $V''(\vq)$ at these Darboux points are
 following
\begin{equation}
\label{eq:li}
 \lambda_1=\frac{1}{c}, \qquad \lambda_2= \frac{2c-1}{1+a^2 +
   \Delta},\qquad \ \lambda_3= \frac{2c-1}{1+a^2 -
   \Delta},
\end{equation}
where 
\[
 \Delta = \sqrt{a^2(2+a^2 -2c)}.
\]
 If potential $V$ is integrable, then by Theorem~\ref{thm:MoRa}, we
 have 
\begin{equation}
\label{eq:2}
\lambda_i\in\scM_{3}=\bigcup_{j=1}^{6}\scM^{(j)}, \mtext{for} 1\leq
i\leq 3,
\end{equation}
where
 \begin{gather*}
   \scM^{(1)}:=\defset{p + \dfrac{3}{2}p(p-1)}{p\in\Z}, \quad 
\scM^{(2)}:=\defset{ \dfrac 1 {2}\left(\dfrac {2} {3}+3p(p+1)\right)}{p\in\Z},\\
   \scM^{(3)}:=\defset{\dfrac 1 {6}\left( 1
       +3p\right)^2-\dfrac 1 {24}}{p\in\Z}, \quad 
\scM^{(4)}:=\defset{\dfrac 3 {32}\left(  1  +4p\right)^2 -\dfrac 1
     {24} }{p\in\Z},\\
\scM^{(5)}:=\defset{ \dfrac 3 {50}\left( 1
      +5p\right)^2-\dfrac 1 {24}}{p\in\Z}, \quad
  \scM^{(6)}:=\defset{\dfrac{3}{50}\left(2 +5p\right)^2-\dfrac1{24}}{p\in\Z}.
 \end{gather*}
From Eq.~\eqref{eq:li} we find that 
\[
c=\dfrac{1}{\lambda_1},\qquad
a=\dfrac{\lambda_1+\lambda_1\lambda_i-2}{\sqrt{2\lambda_1\lambda_i(2-\lambda_1-\lambda_i)}},
\mtext{for} i\in\{2,3\}.
\]
Hence, for arbitrary $\lambda_1,\lambda_2,\lambda_3\in\scM_3$, the
above defined values of $a$ and $c$ give potential~\eqref{eq:vac} which
satisfies the necessary conditions for the integrability.  

Theorem~\ref{thm:MoRa} gives only necessary conditions for the
integrability, and there are many examples that they are not
sufficient. Thus in the above example, we have to check whether
infinitely many potentials are integrable or not.

\section{Integrability of homogeneous polynomial potentials. Global
  analysis}
\label{sec:glob}
In this section we assume that the considered potential $V(\vq)$ is
polynomial and homogeneous of degree $k>2$.  The set of homogeneous
polynomials in $n$ variables $\vq=(q_1,\ldots,q_n)$ of degree $k$, we
denote by $\C_k[\vq]$. 

   During last few years we
worked on the following problem. Is it possible, for a given $k>2$ and
$n>2$, to distinguish all meromorphically  integrable potentials
$V\in\C_k[\vq]$?   In other words, is it possible to give a necessary
and sufficient conditions for the integrability of homogeneous polynomial
potentials? The example given in the end of the previous section shows
that, except Theorem~\ref{thm:MoRa}, we need a result which gives
stronger necessary conditions.    This example shows that even for
fixed $k$ and $n$, Theorem~\ref{thm:MoRa} distinguishes infinitely many
parameters' values for which the potential can be integrable.

\subsection{Darboux points}
\label{sec:dar}
It is clear that the more Darboux points of given potential we know,
the more obstructions for its integrability we obtain from   
the Morales-Ramis Theorem~\ref{thm:MoRa}.  Hence,  we have to know how
many Darboux points a polynomial potential of a given degree can
have.   An analysis of this and similar problems related to
particular solutions of Hamiltonian systems with homogeneous
potentials  forced us  to give a more geometrical definition of Darboux
points.

Let $V$ be a homogeneous polynomial potential of degree $k>2$,
i.e., $V\in\C_k[\vq]$. A direction, i.e., a non-zero $\vd\in\C^n$, is
called \bfi{a Darboux point} of $V$ iff the gradient $V'(\vd)$ of $V$
at $\vd$ is parallel to $\vd$. Hence, $\vd$ is a Darboux point of $V$
iff
\begin{equation}
  \vd\wedge V'(\vd)=\vzero , \qquad \vd\neq \vzero,
\end{equation}
or
\begin{equation}
  V'(\vd)=\gamma \vd, \qquad \vd\neq \vzero,
\end{equation}
for a certain $\gamma\in\C$. Obviously, if $\vd$ satisfies one of the
above conditions, then $\widetilde \vd=\alpha\vd $ for any
$\alpha\in\C^\star$ satisfies them. However, we do not want to
distinguish between $\vd$ and $\widetilde\vd$. Hence we consider a
Darboux point $\vd=(d_1,\ldots, d_n)\in\C^n$ as a point
$[\vd]:=[d_1:\cdots :d_n]$ in the projective space $\CP^{n-1}$.

The set ${\scD}(V) \subset \CP^{n-1}$ of all Darboux points of a
potential $V$ is a projective algebraic set. In fact, ${\scD}(V)$ is
the zero locus in $\CP^{n-1}$ of homogeneous polynomials
$R_{i,j}\in\C_k[\vq]$ which are components of $\vq\wedge V'(\vq)$, i.e.
\begin{equation}
  \label{eq:Rij}
  R_{i,j}:= q_i\pder{V}{q_j}-q_j \pder{V}{q_i}, \mtext{where} 1\leq i<j\leq n.
\end{equation}
We say that a Darboux point $[\vd]\in {\scD}(V)$ is \bfi{a proper
  Darboux point} of $V$, iff $V'(\vd)\neq \vzero$. The set of all
proper Darboux points of $V$ is denoted by ${\scD}^\star(V)$. If
$[\vd]\in {\scD}(V)\setminus {\scD}^\star(V)$, then $[\vd]$ is called
\bfi{an improper Darboux point} of potential $V$.
We say that $[\vd]$ is an \bfi{isotropic Darboux point}, iff
\begin{equation}
  \label{eq:iso}
  d_1^2+\cdots+d_n^2=0.
\end{equation}

We say that
\bfi{potential $V$ is generic} iff all its Darboux points are proper
and simple. The basic fact concerning Darboux points of generic
potentials is given in the following lemma.
\begin{lemma}
  The set of generic potentials $\cG_{n,k}\subset\C_k[\vq]$ of degree
  $k$ is a non-empty open set in $\C_k[\vq]$.  A generic
  $V\in\C_k[\vq]$ has 
\[
D(n,k):=\frac{(k-1)^n - 1}{k-2},
\]
 proper Darboux points.
\end{lemma}
A non-generic potential can have finite, or infinite number of Darboux
points, but for an arbitrary $V\in\C_k[\vq]$ the set $\scD(V)$ is not
empty. Moreover, if $V$ does not have improper Darboux points, then
it has a finite number of proper Darboux points.

\subsection{Obstruction for the integrability due to improper Darboux
  point}
Here we must justify the introduced definition of the Darboux point. Let
us notice that in Theorem~\ref{thm:MoRa} only proper Darboux
points appear and they give particular solutions. However, we have a
more general fact.  
\begin{lemma}
  \label{lem:sls}
  If $[\vd]$ is a proper Darboux point of a homogeneous potential $V$ of
  degree $k>2$, then
  \begin{equation}
    \label{eq:slsp}
    \vq(t):=\varphi(t) \vd,  \quad  \vp(t):=\dot\varphi(t) \vd,
  \end{equation}
  is a solution of Hamilton's equation~\eqref{eq:heqs} provided
  $\ddot \varphi = -\varphi^{k-1}$. Moreover, $V''(\vd)\cdot\vd
  =\lambda_n \vd$ with $\lambda_n=k-1$, and if additionally $[\vd]$ is
  isotropic, then $\lambda_n$ is a multiple eigenvalue of $V''(\vd)$.

  If $[\vd]$ is an improper Darboux point, then
  \begin{equation}
    \label{eq:slsn}
    \vq(t):=t \vd,  \quad  \vp(t):= \vd,
  \end{equation}
  is a solution of Hamilton's equations~\eqref{eq:heqs}. Moreover,
  $V''(\vd)\cdot\vd =\lambda_n\vd$, with $\lambda_n=0$, and if
  additionally $[\vd]$ is isotropic, then $\lambda_n$ is a multiple
  eigenvalue of $V''(\vd)$.
\end{lemma}
Hence,  also an improper Darboux point gives the particular
solution~\eqref{eq:slsn} of the considered canonical
equations~\eqref{eq:heqs}. However, this solution has an extremely
simple form and one can doubt if using it we can obtain any
obstruction for the integrability.  In fact, it is easy to notice that
the monodromy group of the variational equations along
solution~\eqref{eq:slsn} is trivial. Thus, in the frame of the Ziglin
theory we do not obtain any obstacle for the
integrability. Nevertheless, in~\cite{mp:08::e} the following theorem
was proved. 
\begin{theorem}
  \label{thm:im}
  Assume that a homogeneous potential $V\in\C_k[\vq]$ of degree $k>2$
  admits an improper Darboux point $[\vd]\in\CP^{n-1}$. If $V$ is
  integrable with rational first integrals, then matrix $V''(\vd)$ is
  nilpotent, i.e., all its eigenvalues vanish.
\end{theorem}
For $n=2$, the above theorem coincides with  Theorem~2.4
in~\cite{Maciejewski:05::b}. 

\subsection{Relation among eigenvalues}
For a Darboux point $[\vd]\in\scD(V)$ we can calculate eigenvalues
$\lambda_1(\vd), \ldots, \lambda_n(\vd)$ of the Hessian matrix
$V''(\vd)$.  However, numbers $\lambda_i(\vd)$ are not well defined
functions of point $[\vd]\in\CP^{n-1}$, as they depend on its
representative $\vd$. There are several possibilities to define
properly the quantities related to the eigenvalues of $V''(\vd)$ which
do not depend on a choice of a representative of $[\vd]$.  However,
because of some historical reasons and the convention widely accepted
in the literature, we choose the one which is a simple
normalisation. Namely, if $[\vd]$ is a proper Darboux point, then the
chosen representative is such that it satisfies $V'(\vd)=\vd$.
If
$[\vd]$ is an improper Darboux point, then the representative of
$[\vd]$ can be chosen arbitrarily.

Let $[\vd]$ be a proper Darboux point of potential $V$. Then, thanks to
our assumption,  the  eigenvalues $\lambda_1(\vd), \ldots,
\lambda_n(\vd)$ of the Hessian matrix $V''(\vd)$ can be considered as
functions of $\vd$. According to our convention
$\lambda_n(\vd)=k-1$ is the trivial eigenvalue. Let
$\vlambda(\vd)=(\lambda_1(\vd), \ldots, \lambda_{n-1}(\vd))$. Hence we have the
following mapping
\begin{equation}
 \scD^\star(V)\ni [\vd]\longmapsto \vlambda(\vd)\in\C^{n-1}.
\end{equation}
Assume that $\scD^\star(V)$ is finite. Then the image of $\scD^\star(V)$ under
the above map is a finite subset of $\C^{n-1}$. The question is if we can find a
potential $V$ of degree $k$ such that the elements in the image have values
prescribed in advance. We show that the answer to this question is
negative. More precisely, we prove that among $\vlambda(\vd)$ taken at all
proper Darboux points $[\vd]\in\scD^\star(V)$  a certain number of
universal relations exists. These relations play the  central role in our
considerations.

To formulate our first theorem we define  $\vLambda(\vd)=(\Lambda_1(\vd),\ldots,
\Lambda_{n-1}(\vd))$, where $\Lambda_i(\vd):= \lambda_i(\vd)-1$ for $i=1,\ldots,n-1$.
By  $\tau_r$ for $0\leq r \leq n-1$, we denote
the elementary symmetric polynomials in $(n-1)$ variables  of degree $r$, i.e.,
\[
\tau_r(\vx):=\tau_r(x_1,\ldots,x_{n-1})=\sum_{1\leq i_1<\cdots<i_r\leq
n-1}\prod_{s=1}^r x_{i_s}, \qquad 1\leq r\leq n-1,
\]
and $\tau_0(\vx):=1$.

Our first theorem gives the explicit form of the above mentioned relations among
$\vLambda(\vd)$, for a generic potential $V$.
\begin{theorem}
\label{thm:1}
 Let $V\in\C_k[\vq]$ be a homogeneous potential of degree $k>2$ and let all its
Darboux points be proper and simple.
Then
\begin{equation}
 \sum_{[\vd]\in \mathscr{D}(V)}
\frac{\tau_1(\vLambda(\vd))^r}{\tau_{n-1}(\vLambda(\vd)
)}=(-1)^{n-1}(-n-(k-2))^r,
\label{eq:rkoj}
\end{equation}
and
\begin{equation}
 \sum_{[\vd]\in \mathscr{D}(V)}
\frac{\tau_r(\vLambda(\vd))}{\tau_{n-1}(\vLambda(\vd)
)}= (-1)^{r+n-1}\sum_{i=0}^{r}\binom{n-i-1}{r-i}(k-1)^{i},
\label{eq:rtau}
\end{equation}
for $r=0,\ldots,n-1$.
\end{theorem}
Let us explain the importance of Theorem~\ref{thm:1}. To do this we need more
definitions.

 Let $\scC_m$ denote the set of all  unordered tuples  $\vLambda=(\Lambda_1, \ldots,
\Lambda_{m})$, where $\Lambda_i\in\C$ for $i=1,\ldots, m$. For $M>0$, the symbol
$\scC_m^M$ denotes the set of all unordered tuples
$(\vLambda_1,\ldots,\vLambda_M)$, where $\vLambda_i\in\scC_m$, for
$i=1,\ldots,M$.

 We fix $k>2$ and $n\geq 2$, and say that
 a
tuple  $\vLambda\in\scC_{n-1}$ is \bfi{admissible} iff
$\lambda_i=\Lambda_i+1\in\scM_k$ for $i=1,\ldots, n-1$. In other words, $\vLambda_i$ is admissible
iff $\Lambda_i+1$  belongs to
items, appropriate for a given $k$,  in the table of the Morales-Ramis
Theorem~\ref{thm:MoRa}, for $i=1,\ldots, n-1$.
 We denote the set of all
admissible tuples by $\scA_{n,k}$.
If the potential $V$ is
integrable, then for each $ [\vd]\in\scD^\star(V)$, the tuple  $\vLambda(\vd)$ is
admissible.  The set of all admissible elements $\scA_{n,k}$ is countable but infinite.

 If
the set of proper Darboux points of a potential $V$ is non-empty, and $N=\card
\scD^\star(V)$, then the $N$-tuples
\begin{equation}
\label{eq:SV}
\scL(V):=\deftuple{\vLambda(\vd)}{[\vd]\in\scD^\star(V)}\in\scC^N_{n-1},
\end{equation}
is called \bfi{the spectrum of} $V$. Let $\scA^N_{n,k}$ be the subset of
$\scC^N_{n-1}$ consisting of $N$-tuples $(\vLambda_1,\ldots,\vLambda_N)$, such that
$\vLambda_i$ is admissible, i.e., $\vLambda_i\in\scA_{n,k}$, for $i=1,\ldots, N$.
We say that the spectrum $\scL(V)$ of a potential $V$ is admissible iff
$\scL(V)\in\scA^N_{n,k}$. The Morales-Ramis Theorem~\ref{thm:MoRa} says that if
potential $V$ is integrable, then its spectrum $\scL(V)$ is admissible.  However, the problem is that
the set of admissible spectra $\scA_{n,k}^N$ is infinite. We show that from
Theorem~\ref{thm:1} it follows that, in fact, if $V$ is integrable, then its
spectrum $\scL(V)$ belongs to a certain \bfi{finite} subset $\scI^N_{n,k}$ of
$\scA_{n,k}^N$. We call this set  \bfi{distinguished one}, and its elements
\bfi{distinguished spectra}.
\begin{theorem}
\label{thm:ff}
Let potential $V$ satisfy assumptions of Theorem~\ref{thm:1}. If $V$ is
integrable, then there exists a finite subset $\scI_{n,k}^N\subset \scA_{n,k}^N$, where
$N=\card \scD^\star(V)$, such that $\scL(V)\in \scI_{n,k}^N$.
\end{theorem}
Informally speaking, for fixed $k$ and $n$, we restrict the infinite number of
possibilities in each line of the Morales-Ramis table to a finite set of choices.

\subsection{Euler-Jacobi-Kroncker formula and its generalisation}
The importance of Theorem~\ref{thm:ff} is clear.  Having in mind the
our general program of finding all integrable potentials, one would
like to have a generalisations of Theorem~\ref{thm:1} and
Theorem~\ref{thm:ff} for non-generic potentials. In~\cite{mp:07::a} we
gave a proof of  Theorem~\ref{thm:1} using a certain result of
Guillot \cite{Guillot:04::}. Unfortunately,  the methods used in
\cite{Guillot:04::} do not admit such a generalisation.

Our  analysis of case   $n=2$ given in \cite{Maciejewski:05::b} explicitly
showed that one can find an alternative proof of Theorem~\ref{thm:1}
which admits a generalisation to non-generic cases. Moreover it also
gives a clue that an alternative proof~Theorem~\ref{thm:1} can be done
with a help of multidimensional residue technique. We have made many
attempts to find such a proof, however all of them failed. 

Finally we have found amazingly simple solution of the problem. Here
we describe shortly the main construction of our approach. A detailed
exposition the reader will find in \cite{mp:08::e,mp:09::b}. 

Let us introduce local affine coordinates on $\CP^{n-1}$ where the
Darboux points live. We choose chart  $(U_1,\theta_1)$, where 
\[
U_1:=\CP^{n-1}\setminus \defset{[\vq]\in\CP^{n-1}}{q_1 \neq 0},
\]
 and
\begin{equation}
\label{eq:th1}
\theta_1:U_1\rightarrow \C^{n-1}, \quad
 \widetilde\vx:=(x_1,\ldots,x_{n-1})= \theta_1([\vq]),
\end{equation}
 where
\begin{equation}
\label{eq:xi}
x_i=\frac{q_{i+1}}{q_1}, \mtext{for}i=1,\ldots, n-1.
\end{equation}
The image of the set of Darboux points which lie on this chart, i.e., $
\theta_1(\scD(V)\cap U_1)$, is an affine algebraic set 
\begin{equation}
\label{eq:dvu1}
\theta_1(\scD(V)\cap U_1)=\scV(g_1,\dots,g_{n-1}),
\end{equation}
where polynomials $g_1,\ldots,g_{n-1}\in\C[\widetilde \vx]$ are given by
\begin{equation}
\label{eq:g0}
v(\widetilde\vx):=V(1,x_1,\ldots, x_{n-1}), \quad g_0:= k v -\sum_{i=1}^{n-1}
 x_i\pder{v}{x_i},
\end{equation}
and
\begin{equation}
\label{eq:gi}
g_i := \pder{v}{x_i}-x_i g_0, \mtext{for}i=1, \ldots, n-1.
\end{equation}
Moreover, $[\vd]\in \scD(V)\cap U_1$ is an improper Darboux point  iff its affine
coordinates $\widetilde \va := \theta_1([\vd])$ satisfy $g_0( \widetilde
\va)=0$.

It is instructive to consider first the case $n=2$. In this case a Darboux point
$[\vd]$ on the affine chart is given by one coordinate $x_{\star}=\theta_1([\vd])$. It
is a root of polynomial 
\[
g_1(x):=v'(x)-xg_0(x)\mtext{where} g_0(x):=kv(x)-xv'(x).
\]
Moreover, it is easy to notice that the non-trivial eigenvalue
$\lambda(\vd)$ of $V''(\vd)$  can be calculated from the following
formula
\begin{equation}
  \label{eq:lz}
\frac{1}{ \lambda(\vd)-1}=\frac{1}{\Lambda(\vd)}=\frac{g_0(x_{\star})}{g_1'(x_{\star})}.
\end{equation}
The above formula suggests to introduce the following differential
form 
\[
\omega=\frac{g_0(x)}{g_1(x)}\,\rmd x, 
\]
considered as a differential form on $\CP^{1}$. 
This form has poles at  Darboux points. If a Darboux point $[\vd]$ is
proper and simple, then its affine coordinate $x_{\star}$ is a simple
pole of $\omega$, and the residue of $\omega$ at this point is 
\[
\res(\omega,x_{\star})=\frac{g_0(x_{\star})}{g_1'(x_{\star})}=\frac{1}{\Lambda(\vd)}.
\]
Without loss of the generality we can assume that all Darboux points
are located in the affine part of $\CP^{1}$, and then 
\[
\res(\omega,\infty)=1.
\]
Thus, assuming that all Darboux points are proper and simple and
applying the global residue theorem we obtain that 
\[
\sum_{[\vd]\in\scD(V)}\frac{1}{\Lambda(\vd)}=1.
\]
This is just relation~\eqref{eq:rkoj} for $n=2$ with $r=0$. Note that
for $n=2$ it is the only non-trivial relation. 

Now, considering cases with $n>2$ one would like to construct
differential forms in $\CP^{n-1}$ which have poles at Darboux points
and such that their multidimensional
residues at these poles  are given by symmetric functions of
$\vLambda(\vd)$.  It is not difficult to define an
$(n-1)$-differential  form in affine part of $\CP^{n-1}$ which has
poles at Darboux points with residues given in term of symmetric
functions of $\vLambda(\vd)$.  The problem appears with extension of
this form onto whole $\CP^{n-1}$. The obtained global differential
form has some additional, usually not isolated poles and for this type
of forms there is no an appropriate global residue theorem.   

Let us recall basic facts about the multi-dimensional residues and the
Euler-Jacobi-Kronecker formula. For details the reader is refered to
\cite{Aizenberg:83::,Griffiths:76::,Griffiths:78::,Tsikh:92::,Khimshiashvili:06::}.
Let $f_i:\C^n\supset U \rightarrow \C$, where $U$ is an open neighbourhood of
the origin, be holomorphic functions for $i=1, \ldots, n$, and $\vx=\vzero$ be
an isolated common zero of $f_i$. We consider differential $n$-form
\begin{equation}
 \omega:= \frac{p(\vx)}{f_1(\vx)\cdots f_n(\vx)} \,
 \rmd x_1\wedge\cdots\wedge \rmd x_n,
\end{equation}
where $p:U\rightarrow\C$ is a holomorphic function. The residue of the form
$\omega$ at $\vx=\vzero$ can be defined as
\begin{equation}
\label{eq:defres}
 \res(\omega,\vzero):=\frac{1}{(2\pi \rmi)^n}\int_\Gamma \omega,
\end{equation}
where
\begin{equation}
 \Gamma:=\defset{\vx\in U}{ \abs{f_1(\vx)}=
 \varepsilon_1, \ldots,  \abs{f_n(\vx)}=\varepsilon_n},
\end{equation}
and $\varepsilon_1, \ldots, \varepsilon_n$ are sufficiently small positive numbers.
The orientation of $\Gamma$ is fixed by
\begin{equation}
 \rmd(\arg f_1)\wedge\cdots \wedge\rmd(\arg f_n)\geq 0.
\end{equation}
Let us denote $\vf:=(f_1,\ldots, f_n)$. It can be shown that if the Jacobian
$\det \vf'(\vzero)\neq 0$, then
\begin{equation}
\label{eq:0res}
  \res(\omega,\vzero)= \frac{p(\vzero)}{\det \vf'(\vzero)}.
\end{equation}
 The following theorem gives the classical
Euler-Jacobi-Kronecker formula, see e.g. \cite{Griffiths:78::}.
\begin{theorem}[Euler-Jacobi-Kronecker]
\label{thm:ejk}
Let $f_1,\ldots, f_n\in\C[\vx]$ be non-constant polynomials such that
$\scV(\vf):=\scV(f_1,\ldots, f_n)$ is finite and all points of this set are
simple. If $f_1,\ldots, f_n$ do not intersect at the infinity, then for each
$p\in\C[\vx]$ such that
\begin{equation}
\label{eq:degp}
 \deg p\leq \sum_{i=1}^n \deg f_i -(n+1),
\end{equation}
we have
\begin{equation}
\label{eq:ejk}
 \sum_{\vd \in \scV(\vf)}  \res(\omega,\vd)= \sum_{\vd \in \scV(\vf)}  \frac{p(\vd)}{\det \vf'(\vd)} =0.
\end{equation}
\end{theorem}
The above theorem is not sufficient for our investigations. We have to consider
cases when $f_1, \ldots, f_n$ have intersections at the infinity as well as cases when intersections
 of $f_1, \ldots, f_n$ are not simple.

The  homogenisations of $f_i$ are given by
\begin{equation}
 F_i(z_0,z_1,\ldots, z_n):=z_0^{\deg f_i}f_i\left(\frac{z_1}{z_0},\ldots, \frac{z_n}{z_0}\right), \mtext{for}i=1,\ldots, n.
\end{equation}
They define the projective algebraic set $\scV(\vF):=\scV(F_1,\ldots,F_n)\subset
\CP^n$ whose affine part is homeomorphic to $ \scV(\vf)$. Next we extend the
form $\omega$ to a rational form $\Omega$ defined on $\CP^n$. To this end we
consider $\omega$ as the expression of  $\Omega$ on the chart
$(U_0,\theta_0)$. In order to express $\Omega$ on other charts we use the
standard coordinate transformations of  $n$-form. 
Let $[ \vp]=[p_0:\cdots:p_n]\in U_i\cap\scV(F_1,\ldots,F_n)$. We can define
the residue of the form $\Omega$ at point $[\vp]$ as
\begin{equation}
 \res(\Omega, [\vp]):=\res(\widetilde\omega , \theta_i(  [ \vp] )),
\end{equation}
 where $\widetilde\omega$ denotes form $\Omega$ expressed in the chart
 $(U_i,\theta_i)$.

The form $\Omega$ is defined by homogeneous polynomials $F_1,\ldots,F_m$ and
\begin{equation}
 P(z_0,z_1,\ldots, z_n):=z_0^{\deg p}p\left(\frac{z_1}{z_0},\ldots,
 \frac{z_n}{z_0}\right).
\end{equation}
To underline the explicit dependence of $\Omega$ on $F_i$ and $P$ we write
symbolically $\Omega=P/\vF$. The following theorem is a special version of the
global residue theorem.
\begin{theorem}
\label{thm:glo}
 Let $\scV(\vF):=\scV(F_1,\ldots,F_n)$ be a finite set. Then for each polynomial $P$ such that
\begin{equation}
 \deg P\leq \sum_{i=1}^n \deg F_i-(n+1),
\end{equation}
we have
\begin{equation}
 \sum_{[\vs]\in\scV(\vF)} \res(P/\vF, [\vs])=0.
\end{equation}
\end{theorem}
For the proof and the more detailed exposition we refer the reader to \cite{Griffiths:78::,Biernat:92::}.

If $\vzero\in\scV(f)$ is an isolated but not simple point, then we cannot use
formula~\eqref{eq:0res} to calculate the residue of the form $\omega$ at this
point. In such a case we can apply a very nice method developed by Biernat in
\cite{Biernat:89::,Biernat:91::} that reduces the calculation of
multi-dimensional residue to a one dimensional case. We describe it shortly
below.

Let us consider the following analytic set
\begin{equation}
 \scA:=\defset{\vx\in U}{ f_2(\vx)=\cdots=f_n(\vx)=0},
\end{equation}
where $U\subset\C^n$ is a neighbourhood of the origin. Set $\scA$ is a sum of
irreducible one dimensional components $\scA=\scA_1\cup\cdots\cup \scA_m$.
Let $t\mapsto\vvarphi_i(t)\in\scA_i$, $ \vvarphi_i(0)=\vzero$, be an injective
parametrisation of $\scA_i$. Then we define the following forms
\begin{equation}
\label{eq:boi}
 \omega_i = \frac{p(\vvarphi_i(t))}{ \vf'(\vvarphi_i(t))} 
 \frac{f_1'(\vvarphi_i(t))\cdot \dot\vvarphi_i(t)}{f_1(\vvarphi_i(t))}
\rmd t.
\end{equation}
As it was shown in \cite{Biernat:91::} we have
\begin{equation}
 \res(\omega,\vzero)=\sum_{i=1}^m \res(\omega_i, 0).
\end{equation}

In order to use the above theorems we have to make   a kind of
blowup. Roughly speaking, the idea is to associate with a Darboux
point which is located in $\CP^{n-1}$, a finite set of points in
$\CP^n$.

Let us define the following $n$ homogeneous polynomials of $n+1$
variables $\widehat\vq:=(q_0,q_1,\ldots,q_n)$
\begin{equation}
\label{eq:Fi}
F_i:= \pder{V}{q_i}-q_0^{k-2}q_i, \qquad i=1,\ldots, n,
\end{equation}
 and an algebraic set
$\widehat{{\scD}}(V)=\scV(F_1, \ldots, F_n) \subset\CP^n$.

Assume that $[\vd]\in\scD^\star(V)$. Then there exists
$\gamma\in\C^\star$, such that $V'(\vd)=\gamma \vd$, so $k-2$ points
$[\sqrt[k-2]{\gamma}:d_1:\cdots:d_n]\in\CP^n$ belong to
$\widehat{{\scD}}(V)$. These points are well defined as they do not
depend on a representative for $[\vd]$. If $[\vd]$ is an improper
Darboux point, then it defines just one point
$[0:d_1:\cdots:d_n]\in\CP^n$ which is a point of
$\widehat{{\scD}}(V)$.

Set $\widehat{{\scD}}(V)$ is not empty because it contains point
$[\vd_0]:=[1:0:\cdots:0]$. If
$[\widehat\vd]=[d_0:d_1:\cdots:d_n]\in\widehat{{\scD}}(V)\setminus\{[\vd_0]\}$,
then $[\vd]=[d_1:\cdots:d_n]$ is a Darboux point of $V$. Moreover, if $d_0\neq
0$, then $[\vd]$ is a proper Darboux point.

\begin{figure}[h]
\centering \includegraphics*[bb=70.0 442.0 595.0 722.0,scale=0.7]{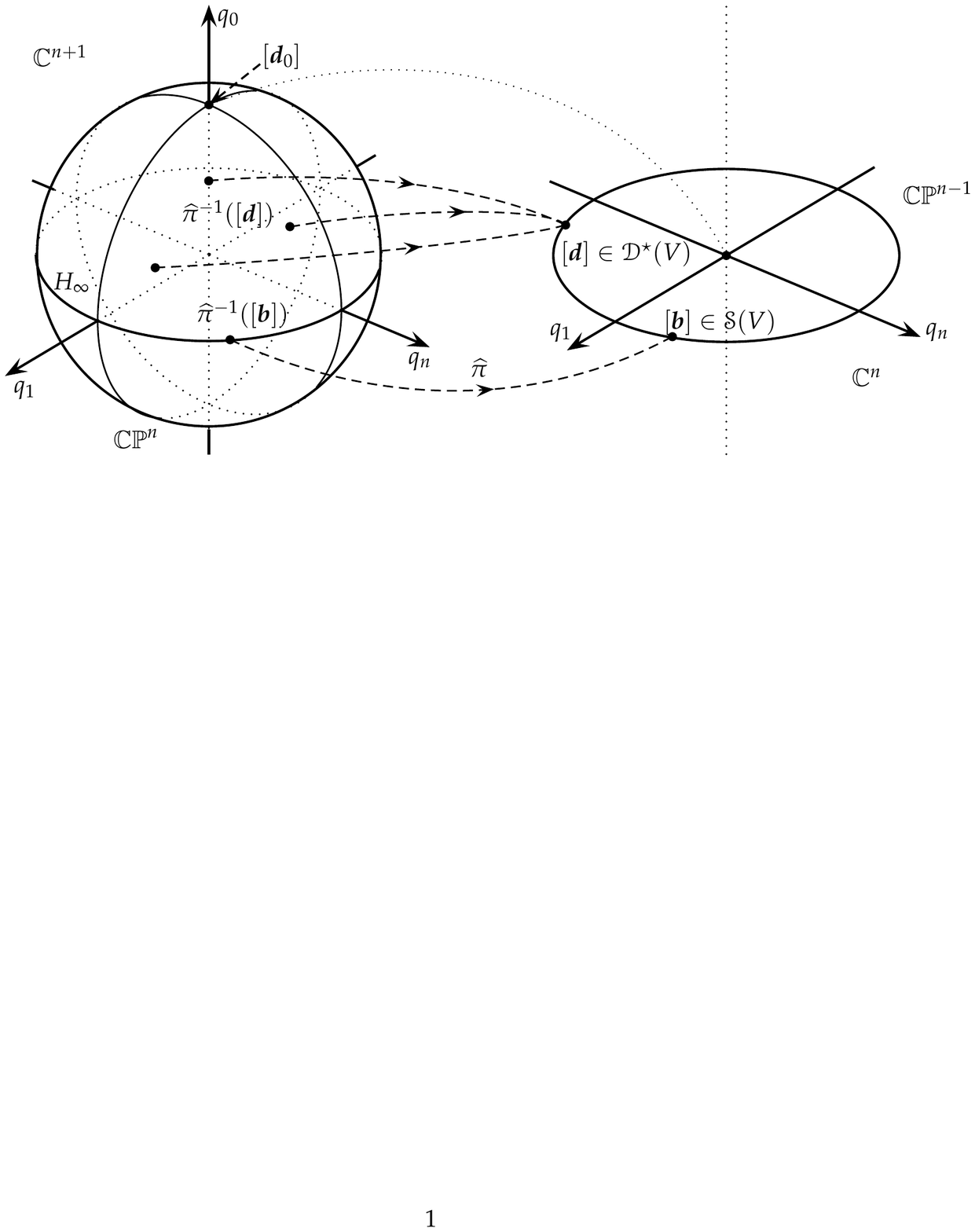}\\[1em]
\caption{Sets $\widehat\scD(V)\subset\CP^n$ and $\scD(V)\subset\CP^{n-1}$}
\label{fig:1}
\end{figure}

The natural projection
\begin{equation}
\label{eq:p}
\pi:\CP^n\setminus\{[\vd_0]\}\rightarrow\CP^{n-1}, \quad
\pi([q_0:q_1:\cdots:q_n])=[q_1:\cdots:q_n],
\end{equation}
maps $\widehat{\scD}(V)\setminus\{[\vd_0]\}$ onto $\scD(V)$, and the intersection
of the inverse image $\pi^{-1}([\vd])$ of a Darboux point $[\vd]\in\scD(V)$ with
$\widehat{\scD}(V)$ is a finite set.  We define also
\begin{equation}
\label{eq:rpi}
\widehat\pi:\widehat{\scD}(V)\setminus\{[\vd_0]\} \rightarrow \scD(V),
\end{equation}
putting $\widehat{\pi}([\widehat{\vd}]):= \pi([\widehat{\vd}])$ for
$[\widehat{\vd}]\in \widehat{\scD}(V)\setminus\{[\vd_0]\}$. That is,
$\widehat{\pi}$ is the restriction of $\pi$ to
$\widehat{\scD}(V)\setminus\{[\vd_0]\}$. This construction is
illustrated in the Figure~\ref{fig:1}.
Now, we can consider differential form $\Omega:=P/\vF$ in $\CP^{n}$,
which in affine part of $\CP^{n}$, is given by 
\[
\omega:=\frac{p_r(\vx)}{f_1(\vx)\cdots f_n(\vx)}, 
\]
where $f_i$ is a dehomogenisation of $F_i$, i.e., 
\begin{equation}
\label{eq:fi}
f_i(\vx):= \pder{V}{x_i}(\vx)-x_i, \mtext{for} i=1,\ldots, n,
\end{equation}
and polynomials $p_r$ are of the form 
\begin{equation}
 p_r(\vx)=(\tr \vf'(\vx) -(k-2))^r, \mtext{with}  r\in\{0,\ldots,
 n-1\}. 
\end{equation}
Now, to obtain relations~\eqref{eq:rkoj} we just make simple
calculations according to the following scheme 
\begin{enumerate}
\item $\scV(\vf):=\scV(f_1,\ldots,f_n)$ is the affine part of
  $\scV(\vF):=\scV(F_1,\ldots,F_n)=\widehat{\scD}(V)$.
\item $\vzero\in \scV(\vf)$, and 
\[
 \vf'(\vzero) =-\vE_n\mtext{so:} p_r(\vzero)=(-n-(k-2))^r, \;\;
\det \vf'(\vzero)=(-1)^n.
\]
\item If $\vd\in\scV(\vf)$, and $\vd\neq\vzero$, then
  $[\vd]\in\scD^\star(V)$ and
\[
 \det\vf'(\vd)=(k-2)\prod_{i=1}^{n-1}\Lambda_i(\vd), \mtext{and} 
p_r(\vd) =\left(\sum_{i=1}^{n-1}\Lambda_i(\vd)\right)^r.
\]
\item For $k>2$, and $n\geq 2$ we have 
\[
\det p_{r}=r(k-2)\leq\sum_{i=1}^{n}\deg f_{i} -(n+1)=  n(k-1)-n -1,
\]
for $r\in \{0, \ldots,n-1\}$.
\item Moreover, $\vd_j:=\varepsilon^j\vd\in\scV(\vf)$, and
  $\vf'(\vd_j)=\vf'(\vd)$ for $j=0, \ldots, k-3$, where $\varepsilon$
  is a primitive $(k-2)$-root of unity.
\item If all Darboux points are proper, then polynomials $f_1, \ldots,
  f_n$ do not intersect at the infinity and we can apply the classical
  Euler-Jacobi-Kronecker formula~\eqref{eq:ejk} from Theorem~\ref{thm:ejk}.
\end{enumerate}
In a case of non-generic potential $V$ having finite number of Darboux
points the potential possesses either improper, or multiple Darboux
points. In such cases we apply Theorem~\ref{thm:glo}. Examples of such
calculations are given in~\cite{mp:09::b}.
\subsection{Applications of global analysis}
In order to perform a reasonable classification of potentials it is
convenient to introduce the following  equivalent relations. 

Let  $\mathrm{PO}(n,\C)$ be the complex projective
orthogonal subgroup of $\mathrm{GL}(n,\C)$, i.e.,
\begin{equation}
\mathrm{PO}(n,\C)=\{\vA\in\mathrm{GL}(n,\C),\ |\ \vA\vA^T=\alpha \vE_n,\;
\alpha\in\C^\star\},
\end{equation}
where $\vE_n$ is $n$-dimensional identity matrix.
We say that $V$ and $\widetilde V$ are equivalent if there exists $\vA\in
\mathrm{PO}(n,\C)$
such that $\widetilde V(\vq)=V_{\vA}(\vq):=V(\vA\vq)$.  Later a
potential means a class of equivalent potentials in the above sense.

The general results described  in the previous section can be applied
to a systematic study the integrability of homogeneous potentials with
fixed $n$ and $k$. 
 The algorithm is following. 

We assume that   $n\geq2$ and $k>2$ are fixed. The aim is to
distinguish all integrable potentials.   

At first  we  consider  generic potentials with $N=  D(n,k)$  proper
Darboux points. 

By Theorem~\ref{thm:ff}, there is only a finite number of distinguished
spectra $\scI_{n,k}^{N}$ and we can find all them  solving Diophantine
equations of the form~\eqref{eq:rkoj} in a subset of rational numbers
defined by the Morales-Ramis table. For example, for $n=2$ we have
$D(2,k)=k$, so a generic homogeneous potential of degree $k$, has $k$ proper
Darboux points. At each proper Darboux point we have one non-trivial
eigenvalue $\Lambda$. Thus, in this case  elements of a
$\scI_{n,k}^{N}=\scI_{2,k}^{k}$, are unordered tuples of $k$
elements.  For $k=3$ and $k=4$ they are listed in Table~\ref{tab:dopuszczalne}

\begin{table}[h]
\begin{center}
\begin{tabular}{l}
\hline
\hline
 $ (\Lambda_1,\Lambda_2, \Lambda_3)$ \\
\hline
\hline
$(-1,-1,1)$ \\
$(-2/3,4,4)$ \\
$(-7/8,14,14)$ \\
$(-2/3,7/3,14)$ \\
\hline
\phantom{123}
\end{tabular}
\hspace*{1cm}
\begin{tabular}{l}
\hline
\hline
  $(\Lambda_1,\Lambda_2, \Lambda_3,\Lambda_4)$ \\
\hline
\hline
$(-1,-1,2,2)$ \\
$(-5/8,5,5,5)$ \\
$(-5/8,2,20,20)$ \\
$(-5/8,27/8,27/8,135)$\\
$(-5/8,2,14,35)$  \\
\hline
\end{tabular}
\end{center}
\caption{\label{tab:dopuszczalne}The distinguished
spectra $\scI_{2,k}^{k}$ for  $k=3$ (left),  and $k=4$ (right) }
\end{table}
It appears that the determination of the distinguished spectra can be
performed only with  a help of a computer algebra system. The known
algorithms used for these purposes  are highly time demanding. 

The next step is to find all possible potentials for a given element
of the distinguished spectrum. In other words we have to determine all
polynomials of a fixed degree such that their Hessians at some points
(which are unknown a priori) have specified eigenvalues. At a first
glance, it seems that this problem is ill-posed. However, it is not
like that. The reason is that the restriction imposed by the fixing of
the eigenvalues is very rigid. Moreover, in fact we work only with
equivalent classes of potentials and this restricts additionally the
number of free parameters in the problem. The algorithm of performing
this step is based on determination of the elimination ideal.  In  all 
cases, for a fixed distinguished spectrum we obtained, either a finite
number of non-equivalent potentials, or a finite number of families of
`separated' potentials (a sum of potentials which depend on smaller
number of variables). 

From the previous step we obtained a finite number of potentials which can be
integrable. To check if they are integrable we use two methods. First
we try to find an additional polynomial first integral applying a direct
method. If it fails we apply the higher order variational equations
in order to prove their non-integrability.   

In a similar way we investigate non-generic cases. 

The first time the prescribed algorithm was used in~\cite{Maciejewski:04::g}
where it was shown that all integrable homogeneous of degree three
polynomial potentials in two variables are already known. Next, in
\cite{Maciejewski:05::b} it was shown for $n=2$ and $k=4$ all
integrable potentials are known except for potential
\[
 V=\dfrac{\alpha}{2}q_1^2(q_1+\rmi q_2)^2+\dfrac{1}{4}(q_1^2+q_2^2)^2,\qquad \alpha\in\C^{\ast},
\]
with $\alpha$ such that  $\lambda=1-\alpha$ belongs to items 2,3 and 5
of table \eqref{eq:tab}.   
All those investigations of homogeneous potentials with two degrees of
freedom allowed to find one new non-trivially integrable potential of
the form 
\begin{equation}
 V_{k,l}=(q_2-\rmi q_1)^l(q_2+\rmi q_1)^{k-l},
\end{equation}
with $k=7$ and $l=2$, which admits an additional polynomial first
integral of degree four in the momenta, see \cite{MR2149780}. 

The case of homogeneous degree three polynomial potentials in three
variables, i.e., case $n=k=3$, was analysed in
\cite{mp:08::e,mp:09::b}.  In this case a generic potential has seven
proper Darboux points $[\vd_{i}]$. At each of them we have pair
$\vLambda(\vd_i):=(\Lambda_1^{(i)}, \Lambda_2^{(i)})$ of the shifted 
eigenvalues.  The relations 
\eqref{eq:rkoj}have the following form
\begin{equation}
 \left.
 \begin{split}
 &\sum_{i=1}^7\frac{1}{\Lambda_1^{(i)}\Lambda_2^{(i)}}=1,\\
&\sum_{i=1}^7\frac{\Lambda_1^{(i)}+\Lambda_2^{(i)}}{\Lambda_1^{(i)}\Lambda_2^{(i)}}= -4,\\
 &\sum_{i=1}^7\frac{(\Lambda_1^{(i)}+\Lambda_2^{(i)})^2}{\Lambda_1^{(i)}\Lambda_2^{(i)}}=16.
 \end{split}
 \quad\right\}
\label{eq:rele3}
\end{equation}
 After really long computations we have found ten distinguished
 spectra. For each of them we reconstructed the general form of the
 potential.  Four among the reconstructed potentials have the form  
\[
 V(q_1,q_2,q_3)=V_1(q_1,q_2) +\dfrac{1}{3}q_3^3,
\]
where  $V_1(q_1,q_2)$ is an integrable potential with two degrees of
freedom. 
Each distinguished spectrum gives several potentials. Some of these
potentials are not integrable, and this fact was proved with the help
of higher order variational equations. 
The remaining ones have the forms 
\[
 \begin{split}
   V_5=&\dfrac{3\rmi }{4}q_1^2q_2 + \dfrac{7\rmi}{3}q_2^3 +
   \dfrac{5}{2}q_2^2q_3 +
   \dfrac{1}{3}q_3^3,\\
   V_6= & 364\sqrt{17}q_1^3 + 2835\rmi\sqrt{17}q_1^2q_2 +
   1560\sqrt{17}q_1q_2^2 +
   6552\rmi \sqrt{17}q_2^3 + \mbox{}\\
   & 4335q_1^2q_3
   + 19074q_2^2q_3 + 578q_3^3,\\
   V_7=& 44 \sqrt{7} q_1^3 + 240\rmi \sqrt{14} q_1^2 q_2 + 330
   \sqrt{7} q_1 q_2^2 + 935\rmi \sqrt{14} q_2^3 +
   3087  q_2^2  q_3 + 294  q_3^3,\\
   V_8=&\dfrac{7}{2}q_1^2q_3-\dfrac{5\rmi\sqrt{3}}{2}q_1^2q_2 -
   \dfrac{9\rmi\sqrt{3}}{2}q_2^3  + \dfrac{15}{2}q_2^2q_3 + \dfrac{1}{3}q_3^3, \\
   V_{9}=&27\rmi \sqrt{3990}q_1^3 + 3726\sqrt{15}q_1^2q_2 -
   456\rmi \sqrt{3990}q_1q_2^2 - 4092\sqrt{15}q_2^3 - \mbox{}\\
   &1125q_1^2q_3 -
   3000q_2^2q_3 - 50q_3^3,\\
   V_{10}=&\dfrac{4\sqrt{2}q_1^3}{3} + \dfrac{5q_1q_2^2}{2\sqrt{2}} +
   q_2^2q_3 + \dfrac{1}{3}q_3^3.
 \end{split}
\]
All the above potentials are integrable. Each of them admits two
commuting additional polynomial first integrals $I_1$ and $I_2$.  
They
were found with the help of a direct method.  All these potentials are
integrable in a non-trivial way, i.e., at least one of additional
first integral is of degree higher than two with respect the momenta.
For example, for the potential 
$V_{10}$ the additional first integrals have the forms 
\[
 \begin{split}
 & I_1=12 p_2^4 - 27 q_2^6 - 18 q_2^4 (q_1^2 - 4  \sqrt{2} q_1 q_3 + 2 q_3^2) +
 4 (6 p_1^2 - 3 p_3^2 + 16  \sqrt{2} q_1^3 - 2 q_3^3) (3 p_3^2 + 2 q_3^3)\\
& +
 12 q_2^2 (3 p_3^2 ( \sqrt{2} q_1 - 4 q_3) + 12 p_1 p_3 (q_1 +  \sqrt{2} q_3) -
   2 q_3^2 (12 q_1^2 +  \sqrt{2} q_1 q_3 + 2 q_3^2))\\
& -
 12 p_2 q_2 (2 p_3 (16 q_1^2 + 3 q_2^2 + 8  \sqrt{2} q_1 q_3 - 4 q_3^2) +
   3  \sqrt{2} p_1 (q_2^2 + 4 q_3^2))\\
& -
 12 p_2^2 (2 p_3 (2  \sqrt{2} p_1 + p_3) - 4 (q_2 - q_3) q_3 (q_2 + q_3) -
    \sqrt{2} q_1 (5 q_2^2 + 8 q_3^2)),
 \end{split}
\]
\[
 \begin{split}
&I_2=81 q_2^8 (2 \sqrt{2} q_1 + q_3) + 216 p_2 p_3 q_2^5 (\sqrt{2} q_1 + 2 q_3)
+
 54 q_2^6 (p_2^2 - 3 p_3^2 + 4 \sqrt{2} q_1^3 - 24 q_1^2 q_3\\
& -
   6 \sqrt{2} q_1 q_3^2) + 384 p_2 p_3 q_1^2 q_2 (3 p_2^2 + 8 \sqrt{2} q_1^3 +
   8 q_1^2 q_3 - 2 \sqrt{2} q_1 q_3^2) - 72 p_1^4 (3 p_3^2 + 2 q_3^3) \\
&+
 144 p_2 p_3 q_2^3 (p_2^2 + 8 q_1^2 (2 \sqrt{2} q_1 + 3 q_3)) +
 144 p_1^3 (\sqrt{2} p_2^2 p_3 + 3 \sqrt{2} p_2 q_2 q_3^2 -
   3 p_3 q_2^2 (q_1 + \sqrt{2} q_3)) \\
&-
 32 (p_2^6 + 12 p_2^4 q_1^2 q_3 + 12 p_2^2 q_1^3 (\sqrt{2} p_3^2 + 4 q_1 q_3^2)
+
   32 q_1^6 (3 p_3^2 + 2 q_3^3)) -
 12 p_1^2 (4 p_2^4 \\
&- 6 p_2 p_3 q_2 (16 q_1^2 + 9 q_2^2 + 8 \sqrt{2} q_1 q_3 -
     4 q_3^2) + 9 q_2^4 (2 q_1^2 + 4 \sqrt{2} q_1 q_3 + q_3^2) +
   32 \sqrt{2} q_1^3 (3 p_3^2 + 2 q_3^3)\\
& +
   12 p_2^2 (p_3^2 + 4 q_2^2 q_3 - \sqrt{2} q_1 (q_2^2 - 2 q_3^2)) +
   6 q_2^2 (9 \sqrt{2} p_3^2 q_1 + 2 q_3^2 (-6 q_1^2 + 2 \sqrt{2} q_1 q_3 +
       q_3^2))) \\
&- 144 q_2^4 (p_2^2 (7 q_1^2 + 5 \sqrt{2} q_1 q_3 + 2 q_3^2) +
   3 q_1^2 (3 p_3^2 - 2 q_3 (-2 q_1^2 + 2 \sqrt{2} q_1 q_3 + q_3^2)))\\
& -
 48 q_2^2 (p_2^4 (5 \sqrt{2} q_1 + 4 q_3) +
   4 p_2^2 q_1^2 (8 q_1^2 + 2 \sqrt{2} q_1 q_3 + 3 q_3^2) +
   8 q_1^3 (9 p_3^2 q_1 + q_3^2 (-6 \sqrt{2} q_1^2\\
& + 4 q_1 q_3 +
       \sqrt{2} q_3^2))) + 6 p_1 (16 \sqrt{2} p_2^4 p_3 +
   16 p_2^2 p_3 (8 q_1^3 - 6 q_1 q_2^2 + 3 \sqrt{2} q_2^2 q_3) +
   4 p_2^3 q_2 (-16 \sqrt{2} q_1^2 \\
&+ 32 q_1 q_3 +
     \sqrt{2} (3 q_2^2 + 4 q_3^2)) +
   3 p_3 q_2^2 (9 \sqrt{2} q_2^4 - 64 q_1^3 (\sqrt{2} q_1 + 2 q_3) -
     12 q_2^2 (2 \sqrt{2} q_1^2 + 8 q_1 q_3\\
& + \sqrt{2} q_3^2)) +
   12 p_2 q_2 (-3 \sqrt{2} p_3^2 q_2^2 + 9 q_1 q_2^4 + 32 q_1^3 q_3^2 +
     4 q_2^2 (4 q_1^3 + 6 q_1 q_3^2 + \sqrt{2} q_3^3))).
 \end{split}
\]
For the potentials $V_7$ and $V_9$ integrals $I_1$ and $I_2$ have, as
above, degrees four and six with respect to the momenta,
respectively. However, they are `monster' first integral---few pages
are necessary to write them down, see \cite{mp:08::e}.

The analysis of non-generic cases is much more involved but,
nevertheless  it can be made almost up to the end, see \cite{mp:09::b}
for details.
\section{Integrability of Newton homogeneous equations}
\label{sec:newtony}

In this section we consider  the following class of Newton's
equations
\begin{equation}
\label{eq:newton}
\ddot \vq= -\vF(\vq), \qquad\vq=(q_1,\ldots,q_n), 
\end{equation}
that we rewrite as  a system of first order differential equations 
\begin{equation}
 \label{eq:newton1}
\dot\vq=\vp, \qquad \dot\vp = -\vF(\vq). 
\end{equation}
Our aim is to investigate  integrability properties  of such
equations. Generally, the forces $\vF$   are not potential, so it can
happen that these equations do not admit even a single first
integral.  

It is easy to observe that equations~\eqref{eq:newton1}  admit
\begin{equation*}
 \mu = \rmd q_1\wedge \cdots \wedge\rmd q_n \wedge \rmd p_1\wedge \cdots
\wedge\rmd p_n,
\end{equation*}
as an invariant $2n$-form.   Thus,  we can talk about  the integrability
in the Jacobi sense of such equations, see
Definition~\ref{def:jac}.  

Now, we want to find  necessary conditions for the
integrability in the Jacobi sense applying the differential Galois
framework.  Thus, we assume just from the beginning that
$(\vq,\vp)\in\C^{2n}$, and we consider only the case when forces are homogeneous
 of degree $(k-1)$. 

 Let us notice formal similarities
between Newton's~\eqref{eq:newton1}, and Hamilton's~\eqref{eq:heqs}
equations.  These similarities allow to define the notion of Darboux
point of a homogeneous force $\vF$, as a non-zero direction  $\vd$
such that $\vF(\vd)$ is parallel to $\vd$.   We say that $\vd$ is a
proper Darboux point iff $\vF(\vd)\neq\vzero$. As in the case of
homogeneous potentials, a Darboux point $\vd$ defines an invariant two
dimensional plane 
\begin{equation}
  \Pi(\vd):= \defset{(\vq,\vp)\in\C^{2n}}{ \vq =\varphi\vd, \  \vp=\psi\vd,
    \quad (\varphi,\psi)\in\C^2 }.
\end{equation}
Equations~\eqref{eq:newton1} restricted to $\Pi(\vd)$ have the form of
one degree of freedom Hamilton's equations
\begin{equation}
  \label{eq:one1}
  \Dt \varphi = \psi, \qquad \Dt \psi = -\varphi^{k-1},
\end{equation}
with the following phase curves
\begin{equation}
  \Gamma_{k,\varepsilon} :=\defset{(\varphi,\psi)\in\C^2}{ \frac{1}{2}\psi^2 +\frac{1}{k}\varphi^k=
\varepsilon}\subset\C^2, \qquad \varepsilon \in\C.
\end{equation}
In this way, a solution $(\varphi,\psi)=(\varphi(t),\psi(t))$
of~\eqref{eq:one1} gives rise a solution
$(\vq(t),\vp(t)):=(\varphi\vd,\psi\vd)$ of equations~\eqref{eq:newton1}
with the corresponding phase curve
\begin{equation}
  \vGamma_{k,\varepsilon} :=\defset{(\vq,\vp)\in\C^{2n}}{ (\vq,\vp)=(\varphi \vd,\psi\vd),\ (\varphi,\psi)\in 
 \Gamma_{k,\varepsilon} }\subset\Pi(\vd).
\end{equation}
The variational equations along $ \vGamma_{k,\varepsilon} $ have the form
\begin{equation}
  \label{eq:var}
   \dot \vx = \vy, \qquad
  \dot \vy =-\varphi(t)^{k-2}\vF'(\vd) \vx,
\end{equation}
or simply
\begin{equation}
 \label{eq:varso}
 \ddot \vx =-\varphi(t)^{k-2}\vF'(\vd) \vx,
\end{equation}
where $\vF'(\vd)$ is the Jacobi matrix of $\vF$ calculated at a Darboux point
$\vd$. Let us
assume that this matrix is diagonalisable.  Then,  in an appropriate basis
equations \eqref{eq:varso} have the form
\begin{equation}
  \label{eq:unve}
  \ddot \eta_i = -\lambda_i \varphi(t)^{k-2}\eta_i,
\qquad i=1,\ldots, n,
\end{equation}
where $\lambda_1, \ldots, \lambda_n$ are eigenvalues of $\vF'(\vd)$.
It is easy to show using the Euler identity that $\vd$ is an
eigenvector of $\vF'(\vd)$ with eigenvalue $k-1$. We always denote it
by $\lambda_n$.  In~\cite{mp:08::a} the following theorem was proved.
\begin{theorem}
 \label{thm:mu}
Assume that the Newton system~\eqref{eq:newton} with  homogeneous
right-hand sides  of degree $(k-1)$ with $k\in\Z^{\star}$, satisfies the following conditions:
\begin{enumerate}
\item force $\vF$ admits a proper  Darboux point $\vd$, and
  $\lambda_1,\ldots,\lambda_n$ are eigenvalues of $\vF'(\vd)$,
\item equations~\eqref{eq:newton1} are integrable in the Jacobi sense
  with first integrals which are meromorphic in a connected
  neighbourhood $U$ of phase curve $\vGamma_{k,\varepsilon}$ with
  $\varepsilon\neq0$, and independent on $U\setminus
  \vGamma_{k,\varepsilon}$.
\end{enumerate}
Then  pairs $(k,\lambda_i)$ for $i=1,\ldots,n$ belong to the
Morales-Ramis Table~\eqref{eq:tab}.
\end{theorem}
The above theorem is a generalisation of two theorems (Theorem~1.2 and
Theorem~1.3) from \cite{mp:08::a}. Here we remark that the similarity
of the theses of Theorem~\ref{thm:MoRa} and Theorem~\ref{thm:mu} is
somewhat misleading. The point is that in the case of Newton equations
we do not have any symplectic structure, and thus we have no all
geometrical consequences of this fact.  On the other hand, the
integrability in the Jacobi sense requires `big' number of first integrals
and exactly this requirement  is the reason why the identity component
of the differential Galois group must be Abelian. 

Fact that we have Theorem~\ref{thm:mu} allows us to think about a
global analysis similar to that we presented in Section~\ref{sec:glob}
for the homogeneous potentials.  In order to analyse this question we
assume the force $\vF$ has polynomial components of the same degree
$l:=k-1>1$, i.e., $\vF\in(\C_l[\vq])^n$.  We denote by $\scD(\vF)$ the
set of all Darboux points of $\vF$. It appears that  a generic force
has $D(n,k)=[(k-1)^n-1]/(k-2)$ Darboux points (considered as points in
$\CP^{n-1}$), i.e., as many as a homogeneous potential $V$ of degree
$k$. It is rather amazing, because $\vF_{V}(\vq):=V'(\vq)$ depends on
$\tbinom{n+k-2}{k-1}$ parameters while $\vF$ depends on
$n\tbinom{n+k-3}{k-2}$ parameters, so in the second case the number of
parameters is much bigger.

For $[\vd]\in\scD(\vF) $, we denote eigenvalues of $\vF'(\vd)$ by
$\lambda_1(\vd),\ldots, \lambda_n(\vd)=(k-1)$, and we set
$\vLambda(\vd):=(\Lambda_1(\vd),\ldots,\Lambda_{n-1}(\vd))$,  where
 $\Lambda_i(\vd):=\lambda_i(\vd)-1$, for $i=1,\ldots, n-1$. 

For the considered Newton equations we can apply all the
methods and tools used for the homogeneous  potentials that are
described in Section~\ref{sec:glob}.  In particular, we have the
following. 
\begin{theorem}
 \label{thm:kojot}
Assume that force $\vF \in(\C_{k-1}[\vq])^n$ has exactly $D(n,k)$
proper  Darboux points $[\vd]\in\scD(\vF)$. Then
$\vLambda(\vd)$ satisfy the following relations:
\begin{equation}
 \label{eq:rkoj1}
 \sum_{[\vd]\in\scD(\vF)}\frac{ \tau_1(\vLambda(\vd))^r
 }{\tau_{n-1}(\vLambda(\vd))}=(-1)^{n-1+r}(n+k-2)^r,
\end{equation}
 or, alternatively
\begin{equation}
 \label{eq:rksym}
 \sum_{[\vd]\in\scD(\vF)}\frac{ \tau_r(\vLambda(\vd))
 }{\tau_{n-1}(\vLambda(\vd))} =
(-1)^{r+1-n}\sum_{i=0}^{r}\binom{n-i-1}{r-i}(k-1)^{i},
\end{equation}
 for $0\leq r\leq n-1$.
\end{theorem}
This theorem has the same consequences for the Jacobi integrability of
homogeneous forces as Theorem~\ref{thm:1} for the integrability of
homogeneous potentials. Namely, for a given $k$, there is only a
finite set of candidates for $\vLambda(\vd)$ satisfying the necessary
conditions for the Jacobi integrability given by
Theorem~\ref{thm:kojot}.  
Hence, we can try to perform an analysis similar to that for
homogeneous potentials, and find all
integrable forces for small $k$. Such an analysis was performed in
\cite{mp:08::a}.  

It appears that for $n=2$ and $k=3$ almost all forces
$\vF\in\C_2[\vq]\times \C_2[\vq]$ are potential. For $k=4$ and
$k=5$ several families of integrable forces were found. Among them some
intriguing non-trivial examples  appear, e.g.,  
force with components   
\begin{equation*}
  F_1 = q_1^2q_2, \quad  F_2=\dfrac{11}{6} q_1 q_2^2.
\end{equation*}
is integrable in the Jacobi sense, with two first integrals which are
both 
of degree four with respect to the velocities. They have the following form
\[
 \begin{split}
 & I_1=24p_1p_2^3+3q_2^2(4q_1^2p_2^2 + 12q_1q_2p_1p_2 -
3q_2^2p_1^2)+16q_1^3q_2^5,\\
&I_2=162p_1^3(q_1p_2  -q_2p_1)-9q_1^3(4q_1^2p_2^2 - 20q_1q_2p_1p_2 +
13q_2^2p_1^2)+
16q_1^6q_2^3.
 \end{split}
\]
It is also worth to mention a remarkable family of forces 
given by 
\begin{equation}
\label{eq:nifam}
 F_1=\lambda q_1 q_2^{k-2}\qquad F_2=q_2^{k-1}, 
\end{equation}
where $k>2$ is and $\lambda\in\C$. The Newton equations with this
force admit  the following  first integral 
\begin{equation*}
 I_1=\frac{1}{2}p_2^2 +\frac{1}{k}q_2^k.
\end{equation*}
If $(k,\lambda)$ belongs to an item of the Morales-Ramis
table~\eqref{eq:tab}, then the system is integrable in the Jacobi
sense with an additional polynomial first integral $I_2$.  Moreover,
for an arbitrary $M>0$ we find $\lambda$ such that the degree of $I_2$
with respect to the momenta is greater than $M$, and there is no an
additional polynomial first integral independent with $I_1$ and having
degree with respect to the momenta smaller or equal to $M$.  Additionally, if
$(k,\lambda)$ belongs to an item different from~2 in
table~\eqref{eq:tab}, then there exist two additional polynomial first
integrals $I_2$ and $I_3$ which are functionally independent together
with $I_1$. The above statements were formulated in a form of a
well justified conjecture  in \cite{mp:08::a}. 

\section{Open problems and perspectives}
\label{sec:open}
Let us consider a two degrees of freedom Hamiltonian system with
homogeneous polynomial potentials of degree $k>2$. In
\cite{Nakagawa:01::} it was shown that if a potential of degree $k>4$
admits a polynomial first integral  of degree higher than two with
respect to the momenta, then this first integral is a product of
polynomial first integrals of lower degrees. One would like to prove
the following.
\begin{conjecture}
  If a homogeneous polynomial potential of degree $k>4$ admits an
  additional polynomial first integral, then it admits an additional
  first integral of degree at most two with respect to the momenta.
\end{conjecture} 
Let us consider a restricted version of this conjecture. Namely, let
us assume that we consider only a generic potential of degree
$k>5$. Such  potential has exactly $k$ proper Darboux points, for
each $k$ we know two elements of its distinguished spectrum
$\scI_{2,k}^{k}$, namely 
\[
\scA_{1,k}=(-1,-1,\underbrace{k-2,\ldots,k-2}_{k-2\ \text{times}}),\qquad
\scA_{2,k}=\Big(-\frac{k+1}{2k},\underbrace{k+1,\ldots,k+1}_{k-1\ \text{times}}\Big).
\]
Let us assume that  for an arbitrary $k>5$  set $\scI_{2,k}^{k}$ has only
these two elements. Then we can prove the restricted version of the
conjecture showing that the only potential with spectrum $ \scA_{1,k}$
is the potential separable in the Cartesian coordinates, and the only
potential with spectrum $\scA_{2,k} $ is a potential separable in
parabolic coordinates. Unfortunately, the assumption about the number
of elements of the set $\scI_{2,k}^{k}$ is true only for $k\leq
13$. For $k=14$, set  $\scI_{2,k}^{k}$ contains two additional
elements 
\[
\begin{split}
\scA_{3,k}=&\Big(-\frac{15}{28},\underbrace{12, \ldots, 12}_{7\
  \text{times}},
\frac{377}{28},\frac{377}{28},15, 15, 780, 5655\Big),\\
\scA_{4,k}=&\Big(-\frac{15}{28},\underbrace{12, \ldots,12}_{8\
  \text{times}},\frac{377}{28},\frac{377}{28},39, 39, 5655\Big).
\end{split}
\]
Then, additional elements of  $\scI_{2,k}^{k}$ appear for $k=17,19, 26, 32, \ldots$, and it
seems that there is no upper bound for these exceptional values of
$k$.

The presented method can be applied effectively only for small values
of $n$ and $k$. With the known computational algorithms for
determination of distinguished spectra and reconstruction of
potential, it seems that limiting values are $n=3$ and $k<5$.

We believe  that  a substantial progress in this field will be
possible after development of  the higher order variational
equations techniques. Till now, they are  used only in cases when the first
order variational equations have a very specific form, see
\cite{Morales:07::}.  

\section*{Acknowledgements}

This research has been partially supported by grant No. N N202 2126 33
of Ministry of Science and Higher Education of Poland and, partially,
by EU funding for the Marie-Curie Research Training Network AstroNet.

%\appendix

%\section{Appendices}

%\section*{References}

% \bibliographystyle{ajmplain}
% \bibliography{mathreva,ajm,yoshida,morales,books,ziglin,churchill,dgt,mp,bogoya1,oldies,audin,moulin,newton,tsygv,ploski}

\newcommand{\noopsort}[1]{}\def\cprime{$'$}
  \def\cydot{\leavevmode\raise.4ex\hbox{.}} \def\cprime{$'$} \def\cprime{$'$}
  \def\cprime{$'$} \def\cprime{$'$} \def\cprime{$'$} \def\cprime{$'$}
  \def\cprime{$'$}

\end{document}